\documentclass{aa}
\pdfoutput=1 

\usepackage{graphicx}
\usepackage{array}
\usepackage{tabularx}
\usepackage{amsmath}
\usepackage{multirow}
\usepackage{txfonts}
\usepackage{natbib}
\bibpunct{(}{)}{;}{a}{}{,} 

\usepackage{grffile} 
\usepackage{floatrow}
\usepackage{caption}
\usepackage{multicol,booktabs}
\usepackage[caption=false]{subfig}

\usepackage{microtype} 

\usepackage{units}

\def\xir{$\xi(r)$}

\def\w0{$w(\theta)$}
\def\wprp{$w_{p}(r_{p})$}
\def\xirppi{$\xi(r_{p},\pi)$ }

\begin{document}

   \title{Evolution of the real-space correlation function from next generation cluster surveys}

   \subtitle{Recovering the real-space correlation function from photometric redshifts}

   \author{Srivatsan Sridhar\inst{1}\thanks{email:ssridhar@oca.eu} \and Sophie Maurogordato\inst{1} \and Christophe Benoist\inst{1} \and Alberto Cappi\inst{1,2} \and Federico Marulli\inst{2,3,4}  }

   \institute{Université C\^{o}te d'Azur, OCA, CNRS, Lagrange, UMR 7293, CS 34229, 06304, Nice Cedex 4, France.
              \email{ssridhar@oca.eu}
         \and
             INAF - Osservatorio Astronomico di Bologna, via Ranzani 1, I-40127, Bologna, Italy. 
         \and
             Dipartimento di Fisica e Astronomia - Universit\`{a} di Bologna,
             viale Berti Pichat 6/2, I-40127 Bologna, Italy
         \and
             INFN - Sezione di Bologna, viale Berti Pichat 6/2, I-40127 Bologna,Italy
         }

   \date{Received XXX; Accepted 01/12/2016}

 
\abstract
   {The next generation of galaxy surveys will provide cluster catalogues probing an unprecedented 
   range of scales, redshifts, and masses with large statistics.
   Their analysis should  
   therefore enable us to probe the spatial distribution of clusters with high accuracy and derive tighter 
   constraints on the cosmological parameters and the dark energy equation of state. However, for 
   the majority of these surveys, redshifts of individual galaxies will be mostly estimated by multiband 
   photometry which implies non-negligible errors in redshift resulting 
   in potential difficulties in recovering the real-space clustering.}
   {In this paper, we investigate to which accuracy it is possible to recover the real-space 
   two-point correlation function of galaxy clusters from cluster catalogues based on photometric 
   redshifts, and test our ability to detect and measure the redshift and mass evolution of 
   the correlation length $r_0$ and of 
   the bias parameter b(M,z) as a function of the uncertainty on the cluster redshift estimate. }
   {We calculate the correlation function for cluster sub-samples covering various mass and 
   redshift bins selected  from a 500 deg$^{2}$ light-cone limited to H$<$24.
   In order to simulate the distribution of clusters in photometric redshift space, 
   we assign to each cluster a redshift randomly extracted from a Gaussian distribution having 
   a mean equal to the cluster cosmological redshift and a dispersion equal to $\sigma_z$. 
   The dispersion is varied in the range $\sigma_{(z=0)}$=$\frac{\sigma_{z}}{1+z_{c}} = 0.005,0.010,0.030$ 
   and $0.050$, in order to cover the typical values expected  in forthcoming surveys.
   The correlation function in real-space is then computed through estimation and deprojection 
   of $w_{p}(r_{p})$. 
   Four mass ranges (from $M_{halo} > 2\times 10^{13}\;h^{-1} M_{\odot}$  to 
   $M_{halo} > 2\times 10^{14}\;h^{-1} M_{\odot}$) and 
   six redshift slices covering the redshift range [0,2] are investigated, first using cosmological 
   redshifts and then for the four photometric redshift configurations.}
   {From the analysis of the light-cone in cosmological redshifts we find a clear increase of the 
   correlation amplitude as a function of redshift and mass. The evolution of the derived bias parameter 
   b(M,z) is in fair agreement with theoretical expectations.
   We calculate the $r_{0}-d$ relation up to our highest mass, highest redshift sample tested 
   ($z=2, M_{halo} > 2\times 10^{14}\;h^{-1} M_{\odot}$).  
   From our pilot sample 
   limited to $M_{halo} > 5\times 10^{13}\;h^{-1} M_{\odot}  (0.4<z<0.7)$, we find that the real-space correlation 
   function can be recovered by deprojection of $w_{p}(r_{p})$ within an accuracy of $5\%$ for 
   $\sigma_{z}=0.001\times(1+z_{c})$ 
   and within $10 \%$ for $\sigma_{z}=0.03\times(1+z_{c})$. For higher dispersions 
   (besides $\sigma_{z}>0.05\times(1+z_{c})$), 
   the recovery becomes noisy and difficult. The evolution of the correlation in redshift and mass is 
   clearly detected for all $\sigma_z$ tested, but requires a large binning in redshift to be detected 
   significantly between individual redshift slices when increasing  $\sigma_z$. The best-fit parameters 
   ($r_0$ and $\gamma$) as well as the bias obtained from the deprojection method for all $\sigma_z$ are 
   within the $1\sigma$ uncertainty of the $z_{c}$ sample.}
   {}


   \maketitle
%


\section{Introduction}
\label{sec:intro}

One of the major challenges in modern cosmology is to explain the observed 
acceleration of the cosmic expansion, determining if it is due to a positive 
cosmological constant, a time--varying dark energy component or a modified 
theory of gravity. Major galaxy surveys are currently ongoing or in preparation
in order to address this fundamental question through the analysis of various 
complementary cosmological probes with different systematics, as for instance 
weak lensing, galaxy clustering (baryon acoustic oscillations, redshift-space
distortions) and galaxy clusters.
In fact galaxy cluster counts as a function of redshift and mass are sensitive 
to dark energy through their dependence on the volume element and
on the structure growth rate. One intrinsic difficulty in constraining the
cosmological models with galaxy cluster counts comes from  
uncertainties in cluster mass estimates and on the difficulty to calibrate related
mass proxies. One can overcome this difficulty 
adding the information related to the clustering properties of clusters, due to
the fact that their power spectrum amplitude depends mainly on the halo mass. 
When combining the redshift-averaged cluster power spectrum and the evolution 
of the number counts in a given survey, the constraints on the dark energy 
equation of state are dramatically improved \citep{majumdar}. Recent 
cosmological forecasting based on galaxy clusters confirms that the figure of 
merit significantly increases when adding cluster clustering information 
\citep{Sartoris_2016}.

Cluster clustering can be measured through the two--point correlation function,
the Fourier transform of the power spectrum, which is one of the most 
successful statistics for analysing clustering processes 
\citep{totsuji_1969,peebles_1980}. In cosmology, it is a 
standard tool to test models of structure formation and evolution.
The cluster correlation function is much higher than that of galaxies, as first
shown by \citet{Bahcall_soneira_d} and \citet{Klypin_1983}.
This is a consequence of the fact that more massive haloes correspond to higher 
and rarer density fluctuations, which have a higher correlation amplitude 
(Kaiser 1984). 
Galaxy clusters are associated to the most massive virialised dark matter 
haloes, and as a consequence their correlation function is strongly amplified.
The evolution of the cluster halo mass, bias and clustering has been addressed 
analytically \citep{Mo_White,Moscardini,Sheth},
and also numerically \citep{Governato_d,Angulo_2005,estrada}.
The increase of the correlation length with cluster mass and 
redshift  has been used to constrain the cosmological model and the bias 
\citep{Colberg_2000,Bahcall_2004,Younger}

The first large local surveys such as the Sloan Digital Sky Survey (SDSS) \citep{SDSS}
have led to significant progress in this field.  
Clustering properties of cluster catalogues derived from SDSS were done by \citet{estrada}, 
\citet{Hutsi} and \citet{Hong_2012}. More recently, 
\citet{Veropalumbo_1} have shown the first unambiguous detection of the Baryon Acoustic Oscillation (BAO) 
peak in a spectroscopic sample of $\sim$25000 clusters selected from the SDSS, and 
measured the peak location at  $104 \pm 7$ $h^{-1}$Mpc . 
Large surveys at higher redshifts which are ongoing such as the
Dark Energy Survey (DES), Baryon Oscillation Spectroscopic Survey (BOSS), Kilo-Degree Survey (KIDS) and  
Panoramic Survey Telescope and Rapid Response System (Pan-STARRS) or in preparation such as the
extended Roentgen Survey with an Imaging Telescope Array (eROSITA), Large Synoptic Survey Telescope (LSST) 
and Euclid open a new window for the analysis of cluster 
clustering. The wide areas covered will give access to unprecedented statistics
($\sim$100,000 clusters expected with Dark Energy Survey, eROSITA and Euclid survey) that will allow us to 
cover the high mass, high redshift tail of the mass distribution, to control 
cosmic variance, and to map the large scales at which the BAO signature is 
expected ($\sim 100$Mpc).

\begin{figure}
\centering
\includegraphics[width=\linewidth]{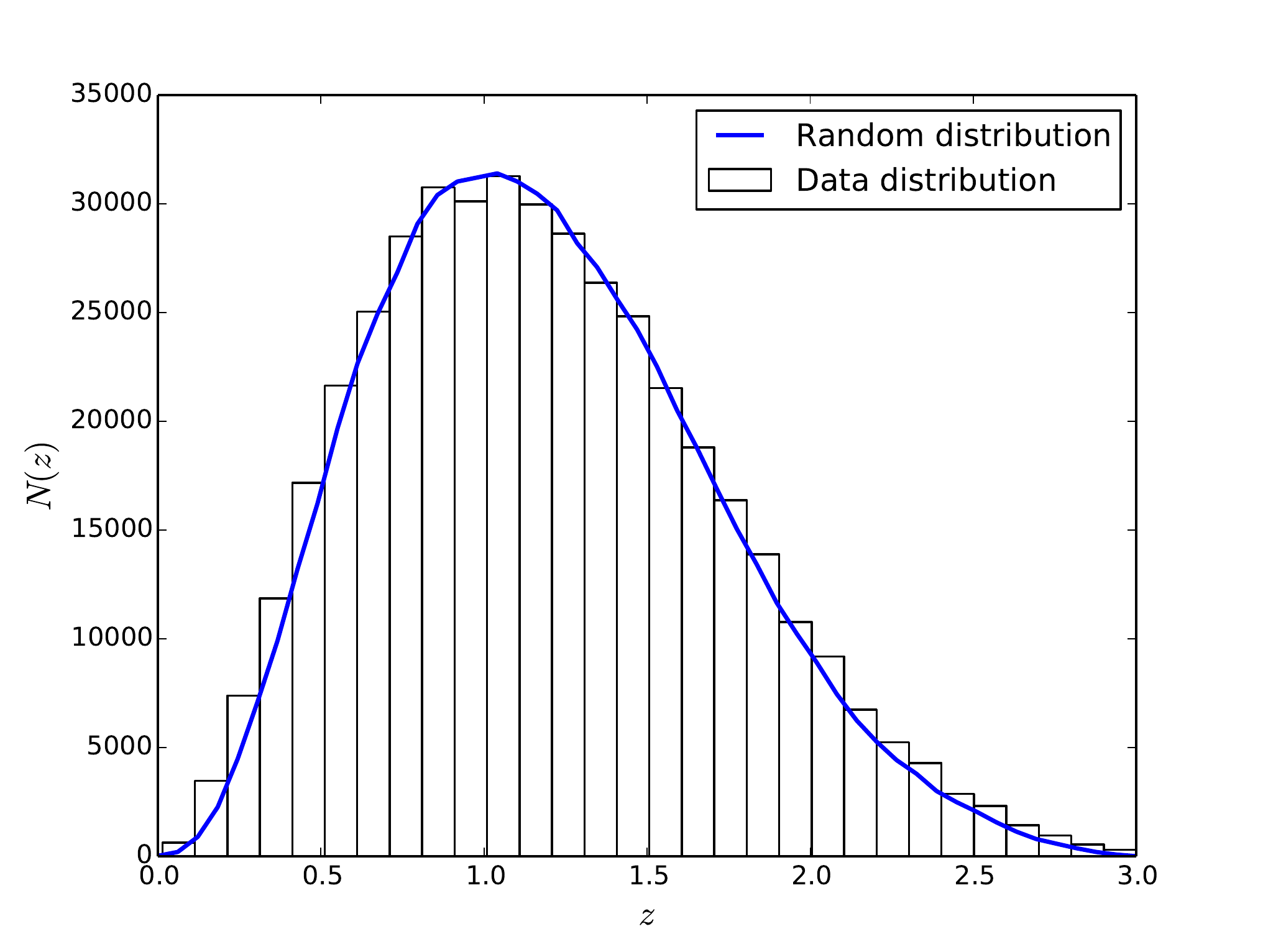}
\caption{Redshift distribution of the entire catalogue with $0.0<z_{c}<3.0$. 
The data distribution is shown as the histogram along with the blue 
line specifying the distribution of the random catalogue we use for calculating
the two-point correlation function.}
\label{fig:hist_random_vs_cluster_dist}
\end{figure}

\begin{figure*}
\subfloat[\label{fig:fits_xir_xidep}]{%
  \includegraphics[height=7.5cm,width=0.49\linewidth]{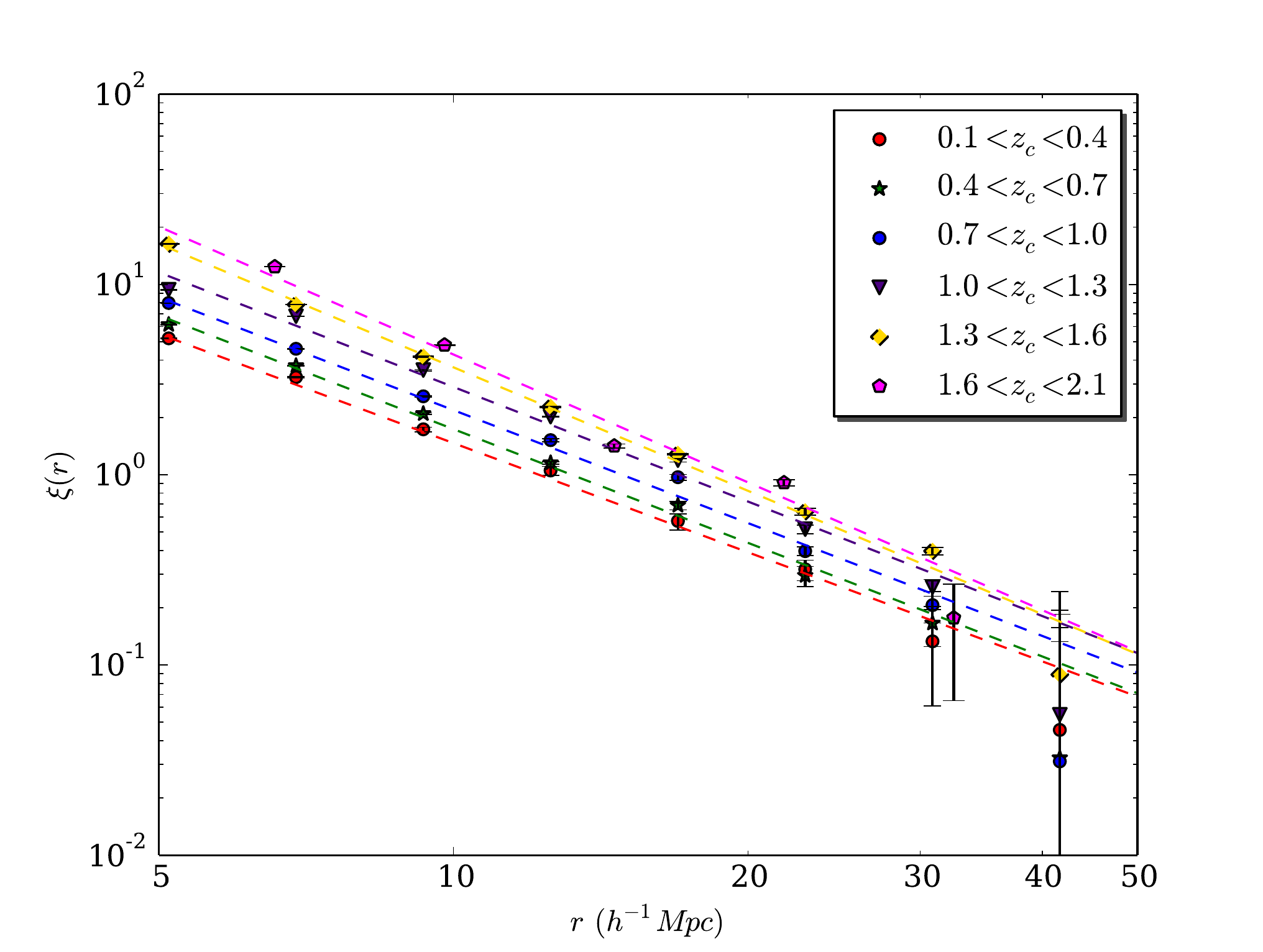}%
}\hfill
\subfloat[\label{fig:fits_xidep}]{%
  \includegraphics[height=7.5cm,width=0.49\linewidth]{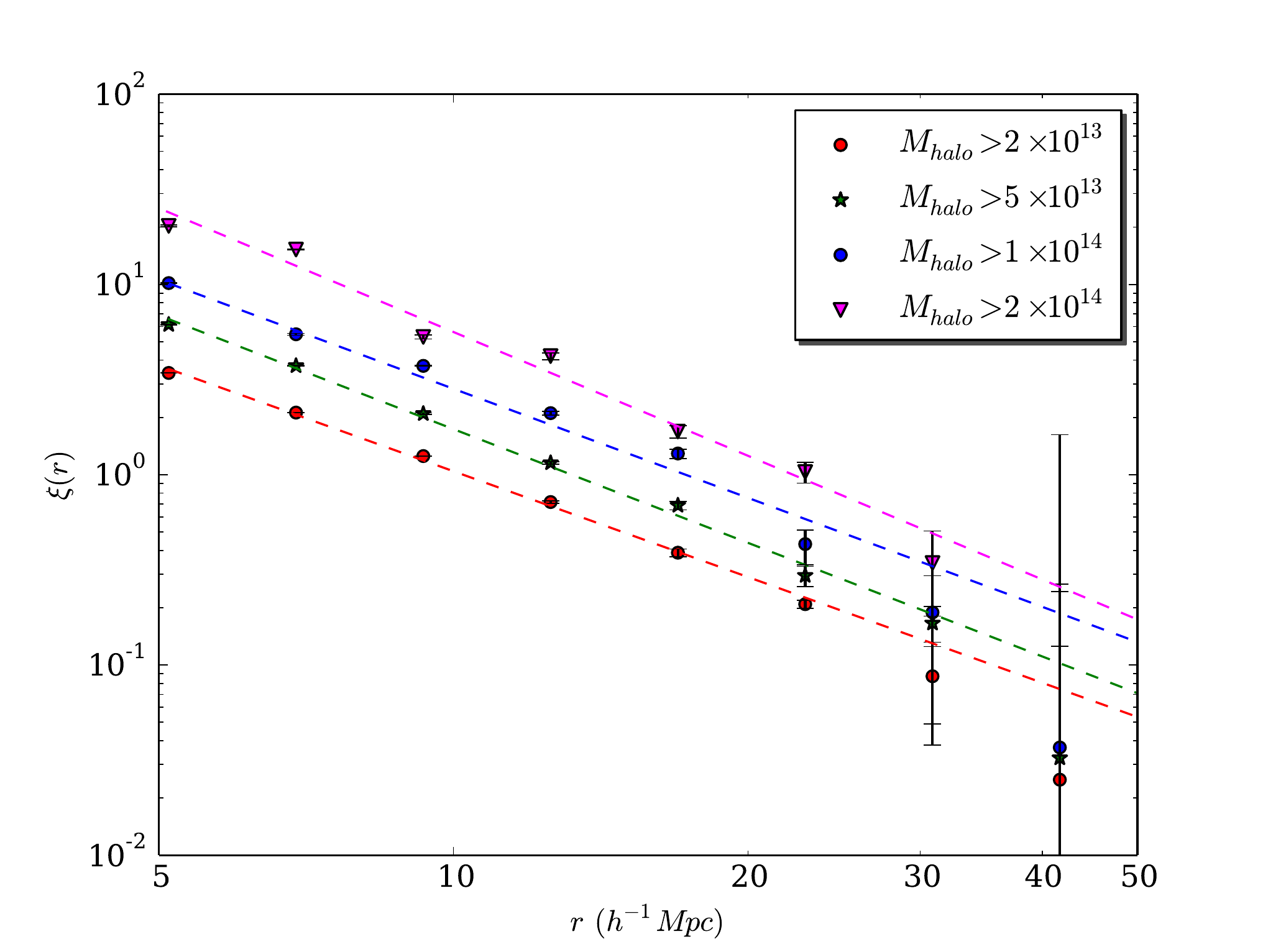}%
}
\caption{Left panel: The correlation functions for clusters with 
$M_{halo}>5\times10^{13}\;h^{-1}\;M_{\odot}$ in six different redshift slices. 
The dashed lines show the corresponding power-law best-fits. The parameter values for the fits can be found 
in Table \ref{table:clustering_with_mass}. Right panel: The correlation functions in the redshift 
slice $0.4<z_c<0.7$ for four different mass cuts (with units $h^{-1}\;M_{\odot}$). 
The dashed lines show the corresponding power-law best-fits. 
The parameter values for the fits can be found in Table \ref{table:clustering_with_mass}. Error bars are 
the square root of the diagonal values of the covariance matrix calculated from the jackknife resampling method.}
\end{figure*}

However, among the several difficulties to be overcome in using clusters as cosmological probes 
is the impact of photometric redshift errors. 
While some of these surveys have (in general partially) a spectroscopic follow 
up, many forthcoming large galaxy surveys will have only the photometric 
information in multiple bands, so that their cluster catalogues will be built 
on the basis of state-of-the-art photometric redshifts. Using those instead of 
real redshifts will cause a positional uncertainty along the line of sight 
inducing a damping of clustering at small scales and a smearing of the acoustic
 peak \citep{estrada}. It is therefore of major interest to check the 
impact of this effect on the recovery of the real-space correlation function. 
Our objective is to optimize the analysis of cluster 
clustering from forthcoming cluster catalogues that will be issued from the 
ongoing and future large multiband photometric surveys. Here we focus on the 
determination of the two-point correlation function, and 
the aim of our paper is i) to determine the 
clustering properties of galaxy clusters from state-of-the-art simulations 
where galaxy properties are derived from semi-analytical modelling \citep{alex_merson}, 
and ii) to test how much the clustering properties evidenced on 
ideal mock catalogues can be recovered when degrading the redshift information to
reproduce the photometric uncertainty on redshift expected in future
cluster experiments.

The paper is organised as follows. In Section \ref{sec:data} we describe the simulation with which we work. 
In Section \ref{sec:two_point_corr} we investigate the two-point correlation function evolution  with redshift 
and mass without any error on the redshift to check consistency with theory. The bias is calculated for 
different mass cut samples along with the evaluation of clustering strength with mass at different redshift 
and is compared with the theoretical expectation of \citet{Tinker_2010}. We also calculate the mean 
intercluster comoving separation ($d$) and compare it with $r_{0}$, and perform an analytic fit to this 
$r_{0}$ vs $d$ relation. Section \ref{sec:section4} presents the deprojection method we use to recover 
the real-space correlation function from mock photometric catalogues generated using a Gaussian 
approximation technique. The results obtained from the deprojection method along with the redshift 
evolution of the samples with redshift uncertainty are presented. 
In  Section \ref{sec:conclusion}, the  overall results obtained from our study are summarised and discussed.


\section{Simulations}\label{sec:data}

We use a public light-cone catalogue constructed using a semi-analytic model of galaxy formation \citep{alex_merson}
onto the $N$-body dark matter halo merger trees of the Millennium Simulation, based on a 
Lambda Cold Dark Matter ($\Lambda$CDM) cosmological model with the following parameters: 
$\Omega_{M},\Omega_{\Lambda},\Omega_{b},h = 0.25, 0.75, 0.045, 0.73$ \citep{springel}, 
corresponding to the first year results from the Wilkinson Microwave Anisotropy Probe \citep{Spergel_2007}. 
The Millennium simulation was carried out using a modified version of the 
GADGET2 code \citep{Springel_from_Merson}. Haloes in the simulation were resolved with a 
minimum of $20$ particles, with a resolution of 
$M_{halo}$ = 1.72$\times 10^{10} h^{-1}M_{\odot}$ ($M_{\odot}$ represents the mass of the Sun). 
The groups of dark matter particles in each snapshot were identified through a 
Friends-Of-Friends algorithm (FOF) following the method introduced by 
\citet{Davis_from_Merson}. 

However, the algorithm was improved with respect to the original FOF,
to avoid those cases where the FOF algorithm
merge groups connected for example by a bridge, while they should be considered instead as 
separated haloes \citep{alex_merson}. 
The linking length parameter for the initial FOF haloes is $b=0.2$. We notice, however, that haloes 
were identified with a method different from the standard FOF. The FOF algorithm was initially applied to find 
the haloes, but then their substructures were identified using the so-called SUBFIND algorithm: depending on 
the evolution of the substructures and their merging, a new final halo catalogue was built. 
The details of this method are described by \citet{Jiang_2014}.

A comparison between the masses obtained with this improved D-TREES algorithm,
$M_{halo}$, and the classical $M_{FOF}$, and their relation with $M_{200}$, was done by \citet{Jiang_2014}, 
where it is shown that 
at redshift $z$ = 0 on average, $M_{halo}$ overestimates $M_{200}$, but by a lower factor with 
respect to $M_{FOF}$: they found that only 5\% of haloes have $M_{halo} /M_{200} > 1.5$. 
However, when comparing the halo mass function of the simulation with that expected from 
the \citet{Tinker_2010} approximation, it appears that there is a dependence on redshift, 
and beyond $z \approx 0.3$ the $M_{halo} /M_{200}$ ratio becomes less than one 
(Mauro Roncarelli, private communication). 
This has to be taken into account in further analysis using the masses. 

Galaxies were introduced in the light-cone using the Lagos 12 GALFORM model \citep{lagos}. 
The GALFORM model populates dark matter haloes with galaxies using a set of differential equations 
to determine how the baryonic components are regulated by "subgrid" physics. These physical processes 
are explained in detail in a series of papers 
\citep{Bower_2006,Font_2008,lagos,alex_merson,Guo_2013,Gonzalez_2014}. 
The area covered by the light-cone is 500 deg$^{2}$; the final mock catalogue is magnitude--limited 
to $H=24$ (to mimic the Euclid completeness) with a maximum redshift at $z = 3$.

For each galaxy the mock catalogue provides different quantities, such as the identifier of the halo in 
which it resides, the magnitude in various passbands, right ascension and declination, and the redshift, 
both cosmological and including peculiar velocities. For each halo in the cluster mass range (see below), 
the redshift was estimated as the mean of the redshifts of its galaxies, while the central right ascension 
and declination were estimated as those of the brightest cluster galaxy (BCG), 
and by construction, the BCG is the centre of mass of the halo.


\section{Evolution of the real-space two-point correlation function in the simulations}\label{sec:two_point_corr}

\subsection{Estimation of the two-point correlation function}\label{subsec:fitting}
In order to measure the clustering properties of a distribution of objects, one of the most commonly used 
quantitative measure is the two-point correlation function \citep{totsuji_1969, davis_peebles_1983}. 
We can express the probability $dP_{12}(r)$  of finding two objects at the infinitesimal volumes $dV_{1}$ 
and $dV_{2}$ separated by a vector distance {\bf r} (assuming homogeneity and isotropy on large scales, 
$r = |{\bf r}|$):

\begin{equation} \label{eqn:prob}
dP_{12} = n^{2}[1+\xi(r)]dV_{1}dV_{2} ,
\end{equation}

where $n$ is the mean number density and the two--point correlation function \xir{} measures the excess 
probability of finding the pair relative to a Poisson distribution. 

Among the various estimators  of the correlation function discussed in the literature 
we use the \citet{landy_szalay} estimator, which has the best performance 
(comparable to the \citet{hamilton} estimator) and is the most popular, being less sensitive to the 
size of the random catalogue and better in handling edge corrections \citep{kerscher}:

\begin{equation}\label{eqn:landy_szalay_est}
\xi(r) = \frac{DD(r) - 2DR(r) + RR(r)}{RR(r)} ,
\end{equation}

where $DD$ is the number of data-data pairs counted within a spherical shell of radius $r$ and $r+dr$, 
$DR$ refers to the number of data-random pairs, and $RR$ refers to the random-random pairs. 

The peculiar motions of galaxies produce redshift-space distortions that have to be taken into account 
in order to recover the real-space clustering \citep{Kaiser_1987}; this means that Equation 
\ref{eqn:landy_szalay_est} cannot be used directly to estimate the 3D real-space correlation function 
when distances are derived from redshifts. We will use it only for the analysis of simulations, where 
the cosmological redshift is available.

The real-space correlation function is expected to follow a power-law as a function of the separation 
$r$ \citep{peebles_1980}:

\begin{equation}\label{eqn:power-law-xir}
\xi(r) = \left(\frac{r}{r_{0}}\right)^{-\gamma} ,
\end{equation} 
where $r_{0}$ is the correlation length and $\gamma$ is the slope. 

The random catalogues we use
reproduce the cluster redshift selection function, estimated by smoothing the cluster redshift distribution 
through a kernel density estimation method. The bandwidth of the kernel is carefully adjusted in 
order to follow the global shape but not the clustering fluctuations in the redshift distribution. 
To ensure that, we use a 
Gaussian kernel two times larger than the bin size, and sample the data in 30 redshift bins. 
Figure \ref{fig:hist_random_vs_cluster_dist} shows the redshift distributions of the simulation 
and of the random catalogue for the whole sample. The random catalogue is ten times denser than 
the simulated sample in order to minimize the effect of shot noise. 

In the following, we estimate the correlation functions for different sub-samples of the original catalogue with 
different cuts in redshift and limiting mass.
Errors are calculated from the covariance matrices using the jackknife resampling method \citep{zehavi_b,norberg}. 
To perform a jackknife estimate we divide the data into $N$ equal sub-samples 
and we calculate the two-point correlation function omitting one sub-sample at a time. 
For $k$ jackknife samples and $i$ bins, the covariance matrix is then given by:

\begin{equation}\label{eqn:jackknife}
C_{ij} = \frac{N-1}{N}\sum_{k=1}^{N}(\xi_{i}^{k}-\bar{\xi_{i}})(\xi_{j}^{k}-\bar{\xi_{j}})
\end{equation} 
where $\bar{\xi_{i}}$ is the average of the values obtained for bin $i$. We make use of $N=9$ 
sub-samples in our calculation. 

To measure the two-point correlation function for all our samples, 
we use \texttt{CosmoBolognaLib}  \citep{CosmoBolognaLib}, a large set of Open Source C++ libraries 
for cosmological calculations. \footnote{More information about \texttt{CosmoBolognaLib} 
can be found at 
\url{http://apps.difa.unibo.it/files/people/federico.marulli3/CosmoBolognaLib/Doc/html/index.html}}


\subsection{Redshift evolution of the cluster correlation function}\label{sec:Redshift_evolution}

\begin{figure}
\centering
\includegraphics[width=\linewidth]{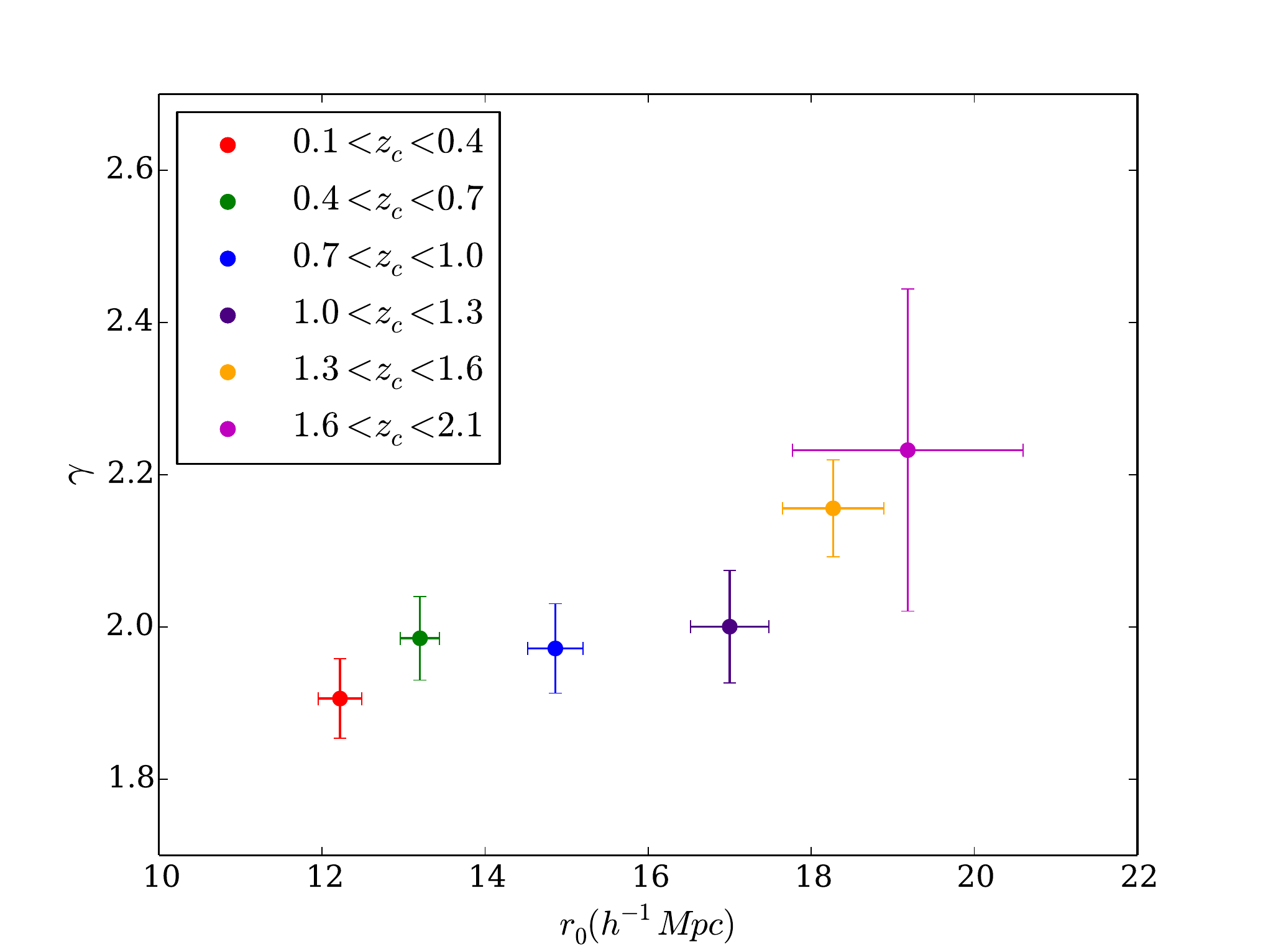}%
\caption{Evolution of $r_{0}$ and $\gamma$ for clusters observed in different redshift slices and with mass $M_{halo}>5\times10^{13}\;h^{-1}\;M_{\odot}$. The values of $r_{0}$ and $\gamma$ can be found in the second panel of Table \ref{table:clustering_with_mass}.}
\label{fig:r0_vs_gamma_zcosmo}
\end{figure}
 
The redshift evolution of the cluster correlation function has been studied both 
observationally \citep{Bahcall_soneira_d, Huchra, Peacock_west, Croft_d, Borgani, Veropalumbo_2}, 
numerically \citep{Bahcall_2004,Younger,Marulli_redshift_evolution} and theoretically 
\citep{Mo_White, Governato_d, Moscardini, Sheth}. Two main results are prominent from these works: 

\begin{itemize}
\item The cluster correlation amplitude increases with redshift for both 
low- and high-mass clusters.
\item The increase of the correlation amplitude with redshift is 
stronger for more massive clusters compared to low-mass ones.
\end{itemize}

Future large surveys are expected to probe 
the high redshift domain with good statistics. This will enable us to study the redshift evolution of clustering 
on a large range of redshifts and  provide independent cosmological tests \citep{Younger}. 
In this section, we investigate the expected redshift evolution of the cluster correlation function in the 
redshift range [0,2], assuming a concordant $\Lambda$CDM model and using the light-cone catalogue detailed 
in Section \ref{sec:data}.

The correlation functions for clusters with a mass cut of $M_{halo}>5\times 10^{13}\;h^{-1}\;M_{\odot}$ 
are estimated in six redshift slices, from $0.1<z_{c}<0.4$ to $1.6<z_{c}<2.1$ 
(where $z_{c}$ refers to the 
cosmological redshift), and are shown in Figure \ref{fig:fits_xir_xidep}.  
The figure shows that, as expected, the amplitude of the cluster correlation function increases with redshift.

For each sub-sample, the correlation function is fitted by a power-law (Equation \ref{eqn:power-law-xir}) 
leaving both $r_{0}$ and $\gamma$ as free parameters. The results of the fits can be visualised in 
Figure \ref{fig:r0_vs_gamma_zcosmo}. 
The redshift range, the values of the best-fit parameters, the number of clusters, 
and the bias (discussed in Section \ref{sec:bias})
for each sub-sample are given in the second panel of Table \ref{table:clustering_with_mass}. 
The fit is performed in the range $5-50 h^{-1}$ Mpc and the error bars are obtained using the 
jackknife estimate method (see Section 
\ref{subsec:fitting}).

The power-law has a relatively stable slope varying between 1.9 and 2.1. In the two highest redshift slices, however, 
the slope appears to be slightly higher, but the variation is at the $\sim 2 \sigma$ level for the $1.3<z_c<1.6$ 
sub-sample and at the $\sim 1 \sigma$ level for the $1.6<z_c<2.1$ sub-sample. On the average, $\gamma \approx 2.0$ 
is close to the measured value for galaxy clusters as done by \citet{totsuji_1969} and \citet{Bahcall_west} on observed
galaxy clusters.

On the contrary, the increase in the correlation length is systematic and statistically significant. When we fix the slope at $\gamma = 2.0$, $r_0$ is shown to increase from $11.97 \pm 0.25$ $h^{-1}$ Mpc for the lowest redshift slice ($0.1<z_{c}<0.4$), to $20.05 \pm 1.13$ $h^{-1}$ Mpc for the highest redshift slice ($1.6<z_{c}<2.1$).
Our results can be compared to \cite{Younger}(see their Figure 5) and are in good agreement with
their analysis of 
the high-resolution simulations of \citet{Hopkins}.

\subsection{The redshift evolution of clustering as a function of mass} 
\label{sec:clustering_with_mass}

\begin{figure}
\centering
\includegraphics[width=\linewidth]{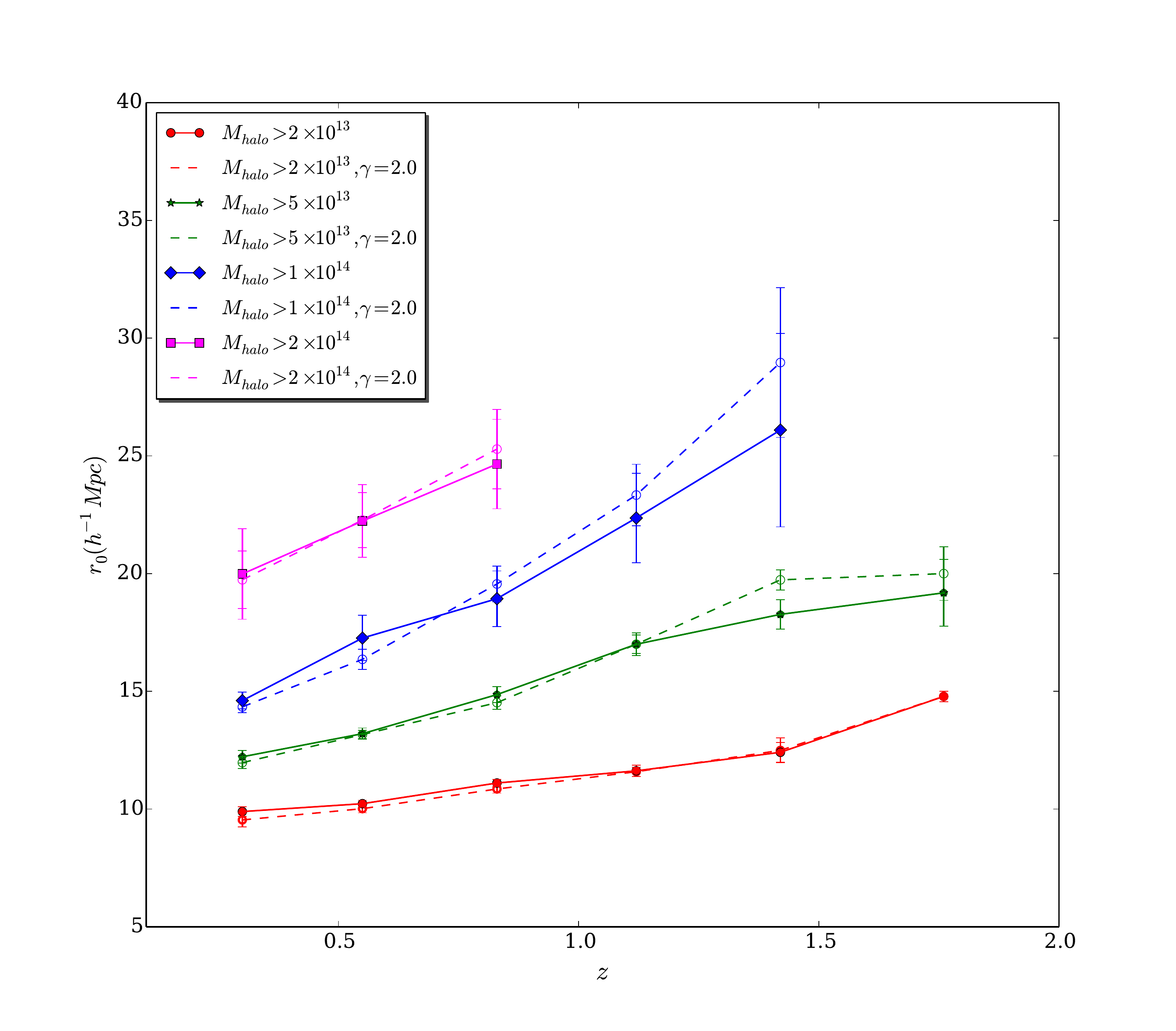}
\caption{Evolution of $r_{0}$ with redshift for different limiting masses (with units $h^{-1}\;M_{\odot}$). 
The filled symbols connected by solid lines correspond to the free slope fits, 
while the open symbols connected by dashed lines correspond to a fixed slope $\gamma = 2.0$.
The different limiting masses are colour coded as shown in the figure. 
The values of $r_{0}$ and $\gamma$ for all the samples can be found in 
Table \ref{table:clustering_with_mass}.}
\label{fig:z_vs_r0_mass_evolution}
\end{figure}

In this section we investigate the redshift evolution of clustering as a 
function of mass. For this purpose, four different mass thresholds are considered: 
$M_{halo}>2\times10^{13}\;h^{-1}\;M_{\odot}$, 
$M_{halo}>5\times10^{13}\;h^{-1}\;M_{\odot}$, 
$M_{halo}>1\times10^{14}\;h^{-1}\;M_{\odot}$ and 
$M_{halo}>2\times10^{14}\;h^{-1}\;M_{\odot}$. 
The analysis is performed in the same redshift slices previously defined. 
The correlation function is fitted with a power-law as it can be seen from Figure \ref{fig:fits_xidep}, 
both with a free slope and with a fixed slope $\gamma = 2.0$. 
The mass range, the values of the best-fit parameters, the number of clusters, 
and the bias for each sub-sample are given in the four panels of Table \ref{table:clustering_with_mass}.  
Each panel corresponds to a different selection in mass. 
In both cases, the correlation length 
$r_{0}$ increases with the limiting mass at any redshift and increases with 
redshift at any limiting mass, as shown in Figure 
\ref{fig:z_vs_r0_mass_evolution}. 
The higher the mass threshold, the larger is the increase of $r_0$ 
with redshift. For example, the ratio of the correlation lengths for the 
[1.3-1.6] and the [0.1-0.4] redshift slices is 1.25 with 
$M_{halo}>2\times10^{13}\;h^{-1}\;M_{\odot}$, 
while it reaches 1.8 with $M_{halo}>1\times10^{14}\;h^{-1}\;M_{\odot}$.
For the largest limiting mass ($M_{halo}>2\times10^{14}\;h^{-1}\;M_{\odot}$), 
the number of clusters becomes small at high $z$ and the analysis must be 
limited to $z=1$.

\begin{figure}
\centering
\includegraphics[width=\linewidth]{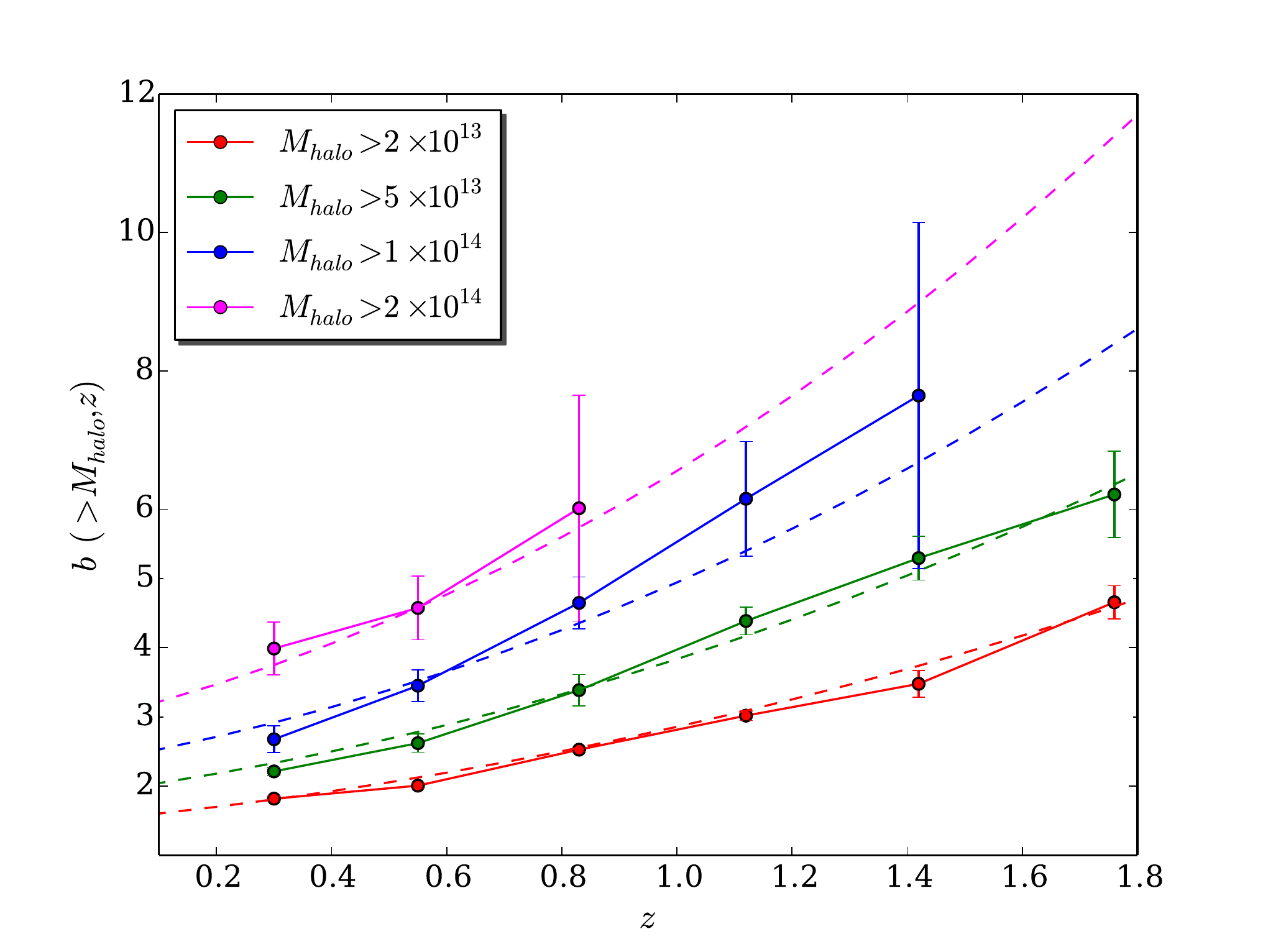}
\caption{Bias as a function of redshift for different limiting masses (with units $h^{-1}\;M_{\odot}$)
where the solid lines just connect the points. The dashed line is the theoretical expectation of the bias
as given by \citet{Tinker_2010} for the same limiting masses and evolving redshift.
The different limiting masses are colour coded as shown in the figure. 
The bias values for all the samples can be found in Table 
\ref{table:clustering_with_mass}. }
\label{fig:z_vs_bias}
\end{figure}

\begin{table*}
\centering
\caption{Best-fit values of the parameters of the real-space correlation function \xir{} for the light-cone at 
different (1) mass thresholds and (2) redshift ranges. For each sample we quote (3) the correlation length $r_{0}$, 
(4) slope $\gamma$, (5) correlation length $r_{0}$ at fixed slope $\gamma=2.0$, (6) number of clusters $N_{clusters}$, 
and (7) the bias $b$ obtained.} 
\label{table:clustering_with_mass}
\begin{tabular}{*{7}{c}} 
\toprule
\multicolumn{1}{p{2cm}}{\centering Mass ($h^{-1}\;M_{\odot}$)}
& \multicolumn{1}{p{2cm}}{\centering $z$}
& \multicolumn{1}{p{2cm}}{\centering $r_{0}$ ($h^{-1}$Mpc)} & 
$\gamma$ & 
\multicolumn{1}{p{2cm}}{\centering $r_{0} \: (\gamma=2.0)$ ($h^{-1}$Mpc)} & $N_{clusters}$ & bias \\ 
\midrule
\multirow{6}{*}{$M_{halo}>2\times10^{13}$} & $0.1<z_c<0.4$ & 9.89$\pm$0.20 & 1.76$\pm$0.05
			 & 9.53$\pm$0.29 & 10492 & 1.81$\pm$0.03 \\ [1ex] 
   		& $0.4<z_c<0.7$ & 10.22$\pm$0.14 & 1.84$\pm$0.04 & 10.01$\pm$0.17 & 27224 & 2.00$\pm$0.03 \\ [1ex] 
   		& $0.7<z_c<1.0$ & 11.10$\pm$0.15 & 1.87$\pm$0.04 & 10.85$\pm$0.17 & 35133 & 2.52$\pm$0.02 \\ [1ex] 
   		& $1.0<z_c<1.3$ & 11.62$\pm$0.23 & 1.98$\pm$0.05 & 11.58$\pm$0.19 & 31815 & 3.01$\pm$0.06 \\ [1ex] 
   		& $1.3<z_c<1.6$ & 12.41$\pm$0.42 & 2.13$\pm$0.09 & 12.49$\pm$0.52 & 22978 & 3.37$\pm$0.19 \\ [1ex] 
   		& $1.6<z_c<2.1$ & 14.78$\pm$0.21 & 2.06$\pm$0.05 & 14.78$\pm$0.22 & 18931 & 4.65$\pm$0.23 \\ [1ex] \hline \noalign{\vskip 0.1cm}    
\multirow{6}{*}{$M_{halo}>5\times10^{13}$} & $0.1<z_c<0.4$ & 12.22$\pm$0.26 & 1.90$\pm$0.05
			 & 11.97$\pm$0.25 & 3210 & 2.21$\pm$0.05 \\ [1ex] 
   		& $0.4<z_c<0.7$ & 13.20$\pm$0.23 & 1.98$\pm$0.05 & 13.16$\pm$0.17 & 7301 & 2.62$\pm$0.13 \\ [1ex] 
   		& $0.7<z_c<1.0$ & 14.86$\pm$0.33 & 1.97$\pm$0.05 & 14.52$\pm$0.28 & 8128 & 3.38$\pm$0.22 \\ [1ex] 
   		& $1.0<z_c<1.3$ & 17.00$\pm$0.48 & 2.00$\pm$0.07 & 17.00$\pm$0.38 & 5963 & 4.38$\pm$0.19 \\ [1ex] 
   		& $1.3<z_c<1.6$ & 18.26$\pm$0.62 & 2.15$\pm$0.06 & 19.73$\pm$0.43 & 3365 & 5.29$\pm$0.31 \\ [1ex] 
   		& $1.6<z_c<2.1$ & 19.18$\pm$1.41 & 2.23$\pm$0.21 & 20.05$\pm$1.13 & 2258 & 6.21$\pm$0.62 \\ [1ex] \hline \noalign{\vskip 0.1cm}    
\multirow{5}{*}{$M_{halo}>1\times10^{14}$} & $0.1<z_c<0.4$ & 14.60$\pm$0.35 & 1.93$\pm$0.06
			 & 14.33$\pm$0.24 & 1119 & 2.67$\pm$0.19 \\ [1ex] 
   		& $0.4<z_c<0.7$ & 17.26$\pm$0.96 & 1.90$\pm$0.08 & 16.35$\pm$0.42 & 2228 & 3.45$\pm$0.23 \\ [1ex] 
   		& $0.7<z_c<1.0$ & 18.93$\pm$1.18 & 2.08$\pm$0.12 & 19.55$\pm$0.75 & 2072  & 4.64$\pm$0.37 \\ [1ex] 
   		& $1.0<z_c<1.3$ & 22.36$\pm$1.90 & 2.11$\pm$0.17 & 23.33$\pm$1.30 & 1221  & 6.15$\pm$0.82 \\ [1ex] 
   		& $1.3<z_c<1.6$ & 26.09$\pm$4.10 & 2.28$\pm$0.30 & 28.96$\pm$3.17 & 590 & 7.64$\pm$2.50 \\ [1ex] \hline \noalign{\vskip 0.1cm}    
\multirow{3}{*}{$M_{halo}>2\times10^{14}$} & $0.1<z_c<0.4$ & 19.98$\pm$1.92 & 1.95$\pm$0.22
			 & 19.73$\pm$1.22 & 322 & 3.98$\pm$0.38 \\ [1ex] 
   		& $0.4<z_c<0.7$ & 22.23$\pm$1.54 & 2.16$\pm$0.18 & 22.27$\pm$1.17 & 538 & 4.57$\pm$0.45 \\ [1ex] 
   		& $0.7<z_c<1.0$ & 24.65$\pm$1.89 & 2.19$\pm$0.29 & 25.28$\pm$1.68 & 407 & 6.01$\pm$1.63 \\ [1ex]  \hline  \\    
\end{tabular}
\end{table*}

We can again compare our results with the analysis of \citet{Younger}.
who used a Tree Particle Mesh (TPM) code \citep{Bode_Ostriker} to 
evolve $N$ = 1260$^{3}$ particles in a box of 1500 $h^{-1}$Mpc, reaching a redshift $z \approx 3.0$. 
We find a good agreement (see their figure 5)
for the common masses and redshift ranges tested; our analysis probes the correlation function of 
$M_{halo}>1\times10^{14}\;h^{-1}\;M_{\odot}$ clusters up to 
$z \approx 1.6$ , and of
$M_{halo}>2\times10^{14}\;h^{-1}\;M_{\odot}$ clusters up to $z \approx 0.8$,
thus extending the $r_0(z)$ evolution shown by \citet{Younger} 
to higher redshifts for these high mass clusters.


\subsection{Bias evolution}\label{sec:bias}

Starting from the initial matter density fluctuations, structures grow 
with time under the effect of gravity. The distribution of haloes, 
and hence of galaxies and clusters, is biased with respect to the 
underlying matter distribution, and on large scales it is expected that the 
bias is linear:

\begin{subequations}
\begin{equation}
\left( \dfrac{\Delta \rho}{\rho}\right)_{light} = 
b \times \left( \dfrac{\Delta \rho}{\rho}\right)_{mass} ,
\end{equation} 

where $b$ is the bias factor and $\rho$ is the density.
The higher the halo mass, the higher the bias.

The amplitude of the halo correlation function is
amplified by a $b^2$ factor with respect to the matter correlation function:

\begin{equation} \label{eqn:bias}
\xi(r)_{light} = b^{2} \times \xi(r)_{mass} .
\end{equation}
\end{subequations}

The amplitude of the matter correlation function increases with time and 
decreases with redshift, but the halo bias decreases with time and 
increases with  redshift. As a result, the cluster correlation amplitude 
increases with redshift, as shown clearly in Figure \ref{fig:r0_vs_gamma_zcosmo}.

We estimate the cluster bias through Equation \ref{eqn:bias}.
The power spectrum of the dark matter distribution is calculated 
with the cosmological parameters of the light-cone we use and 
we obtain its Fourier transform \xir{}. We use the function 
\texttt{xi\_DM} from the class \texttt{Cosmology} 
from \texttt{CosmoBolognaLib}. 
The comparison is not straightforward because, as we have previously noticed, 
in the simulation halo masses are
not $M_{200}$, but the so--called Dhalos masses $M_{halo}$.

The evolution of bias with redshift for the four sub-samples with different
limiting masses is shown in Figure \ref{fig:z_vs_bias} along with the theoretical predictions of 
\citet{Tinker_2010} for the same limiting masses used. The values of the bias for the different sub-samples
are provided in Table \ref{table:clustering_with_mass}. We can see that at fixed redshifts 
more massive clusters have a higher bias; at fixed mass threshold the bias increases with 
redshift, and evolves faster at higher redshifts. It can also be seen clearly that the bias 
obtained for the haloes from the simulations are in good agreement with the predictions by
\citet{Tinker_2010}.

At high redshifts ($z>0.8$) the bias recovered from the simulations seems to slightly 
diverge from the theoretical predictions, 
especially for the $M_{halo}>1\times10^{14}\;h^{-1}\;M_{\odot}$ sample, but this can be explained by the dependence on redshift for the $M_{halo}/M_{200}$ ratio which  becomes 
smaller than one at these redshifts as previously mentioned. 
The discrepancy is not significant as our bias measurements are 
well within $1\sigma$ precision from the theoretical expectations.


\subsection{The $r_{0}$ - \textit{d} relation } 
\label{sec:r0_vs_d_relation}

The dependence of the bias on the cluster mass is based on theory. 
A complementary and empirical characterization of the cluster 
correlation function is the dependence of the correlation length $r_{0}$  
as a function of the mean cluster comoving 
separation $d$ \citep{Bahcall_soneira_d,Croft_d,Governato_d,Bahcall_2003}, 
where:

\begin{equation}
d = \sqrt[3]{\tfrac{1}{\rho}}
\end{equation}

and  $\rho $ is the mean number density of the cluster catalogue on a given 
mass threshold. 

According to the theory, more massive clusters have a higher bias, therefore a 
higher $r_0$; as they are also more rare, they have also a larger mean 
separation: therefore it is expected that $r_0$ increases with $d$, 
that is, $r_0 = \, \alpha \, d^{\,\beta}$.

This relation has been investigated both in observational data 
\citep{Bahcall_west,estrada} and in numerical simulations 
\citep{Bahcall_2003, Younger}. \citet{Younger} gave an analytic 
approximation in the $\Lambda$CDM case in the redshift range $z$ = 0-0.3 for 
$20\leq d \leq 60 \; h^{-1}$Mpc, with $\alpha = 1.7$ and $\beta = 0.6$.

\begin{figure}
\centering
\includegraphics[width=\linewidth]{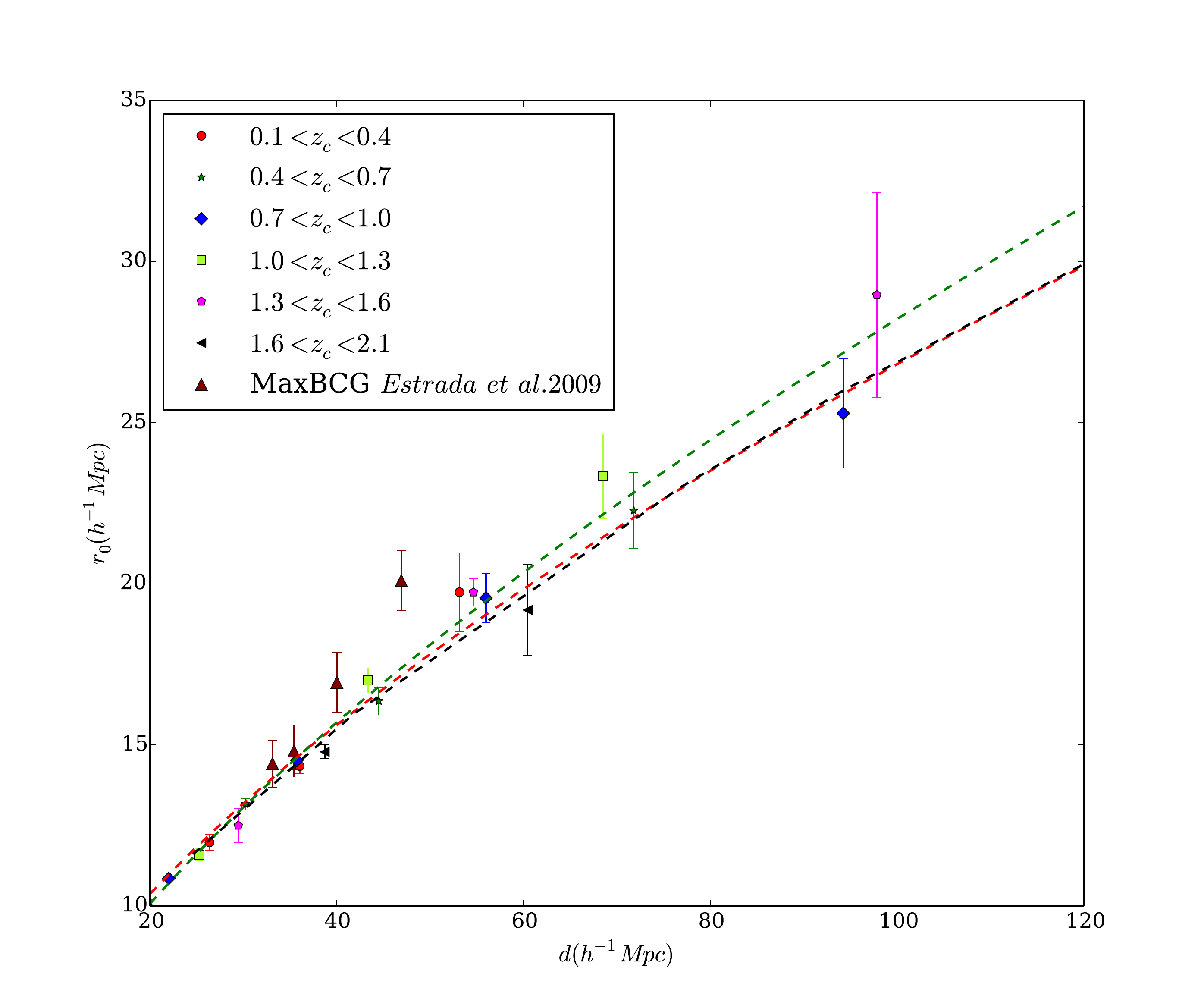}
\caption{Evolution of $r_{0}$ with $d$ for clusters 
of different masses in different redshift slices. The points plotted are for the fixed slope $\gamma=2.0$.
The green dashed line shows the fit when $\gamma=2.0$ and 
the red dashed line shows the overall fit obtained for the data points considering a free slope. 
The analytic approximation in the $\Lambda$CDM case obtained by \citet{Younger} 
is shown by the dashed black line. The different redshift slices are colour coded as mentioned 
in the figure. }
\label{fig:d_vs_r0_relation}
\end{figure}

We determine the $r_{0}$ dependence with $d$ for the various sub-samples 
previously defined.
The results obtained for a free $\gamma$ along with the best fit obtained for 
both free and fixed $\gamma=2$ are shown 
in Figure \ref{fig:d_vs_r0_relation}. 
The best-fit parameters for the $r_{0}$ - $d$  relation in the redshift range $0 \le z \le 2.1$ 
and for the cluster mean separation range 
$20\leq d \leq \; 140 h^{-1}$Mpc are, $\alpha = 1.77\pm0.08$ and $\beta = 0.58\pm0.01$.
The $r_{0}$ - $d$  relation which appears to be scale--invariant with 
redshift, is consistent with what was found by \citet{Younger} and is also consistent 
with the theoretical predictions of \citet{estrada} (see their figure 7).

The scale invariance of the $r_{0}-d$ relation up to a redshift 
$z \approx 2.0$ implies that the increase of the cluster correlation strength
 with redshift is matched by the increase of the mean cluster separation 
$d$. It suggests that the 
cluster mass hierarchy does not evolve significantly in the tested redshift 
range: for example, the most massive clusters at an earlier epoch will still 
be among the most massive at the current epoch.

\section{Estimating the correlation function with photometric redshifts}
\label{sec:section4}

Large redshift surveys such as SDSS \citep{SDSS}, 
VIMOS VLT Deep Survey (VVDS) \citep{VVDS}, VIMOS Public Extragalactic Redshift Survey (VIPERS) 
\citep{VIPERS_1,VIPERS_2} have revolutionised our tridimensional vision of the Universe. 
However, as spectroscopic follow-up is a very time consuming task, priority has been given either to the
sky coverage or to the depth of the survey. 
An alternative way of recovering the redshift information is to derive it from imaging in multiple bands 
when available, using the technique of photometry \citep{Ilbert_2006,Ilbert_2009}.
While the accuracy of spectroscopic redshifts cannot be reached, this method can be successfully used for 
several purposes such as, for instance, cluster detection.
Several major surveys that will provide imaging in multiple bands and thereby photometric redshifts 
are in progress or in preparation. For instance, the ongoing Dark Energy Survey aims to 
cover about 5000 deg$^{2}$ of the sky  with a photometric accuracy of 
$\sigma_{z} \approx 0.08$ out to $z\approx1$ \citep{Sanchez_DES}. 
Future surveys such as LSST \citep{LSST_1,LSST_2} and Euclid 
\citep{euclid_red_book} are expected to make a significant leap forward.  
For instance, the Euclid Wide Survey, planned to cover 15000 deg$^{2}$, 
is expected to deliver photometric redshifts with uncertainties lower than $\sigma_{z}/(1+z) < 0.05$ 
(and possibly  $\sigma_{z}/(1+z) < 0.03$) \citep{euclid_red_book} over the  redshift range [0,2].
The performances of photometric redshift measurements have significantly increased over the last decade, 
making it possible to perform different kinds of
clustering analysis which were previously the exclusive domain of
spectroscopic surveys. 
In this section we investigate how well we can recover the cluster correlation 
function for a sample of clusters with photometric redshifts and test the impact of
the photometric redshift errors in redshift and mass bins.


\subsection{Generation of the photometric redshift distribution of haloes}
\label{sec:mock_photo_samples}

From the original light-cone we extracted 
mock cluster samples with photometric redshifts; these were assigned to each 
cluster by random extraction from a Gaussian distribution with mean equal 
to the cluster cosmological redshift and standard deviation equal to the
assumed photometric redshift error of the sample.

In this way we built five mock samples with errors
$\sigma_{(z=0)} = \sigma_{z}/(1+z_{c}) = 0.001, 0.005, 0.010, 0.030, 0.050$.
These values have been chosen to span the typical uncertainties 
expected in the context of upcoming large surveys.

The photometric redshift uncertainties in upcoming surveys are expected to be 
within 0.03$< \sigma_{z}/(1+z) <$0.05 for galaxies and  within 0.01$< \sigma_{z}/(1+z) <$0.03 
for clusters \citep{Ascaso}. 
Ideally, the error on the cluster redshift should scale proportionally to 
${N_{mem}}^{-1/2}$ (where $N_{mem}$ is the number of cluster members), therefore for clusters with ten detected members
the error would be reduced by a factor three; but of course
contamination from non-member galaxies will affect the redshift estimate.

\begin{figure*}
	\centering
     \setlength{\unitlength}{1cm}
     \begin{picture}(16.0, 14.2)
     \includegraphics[scale=0.45]{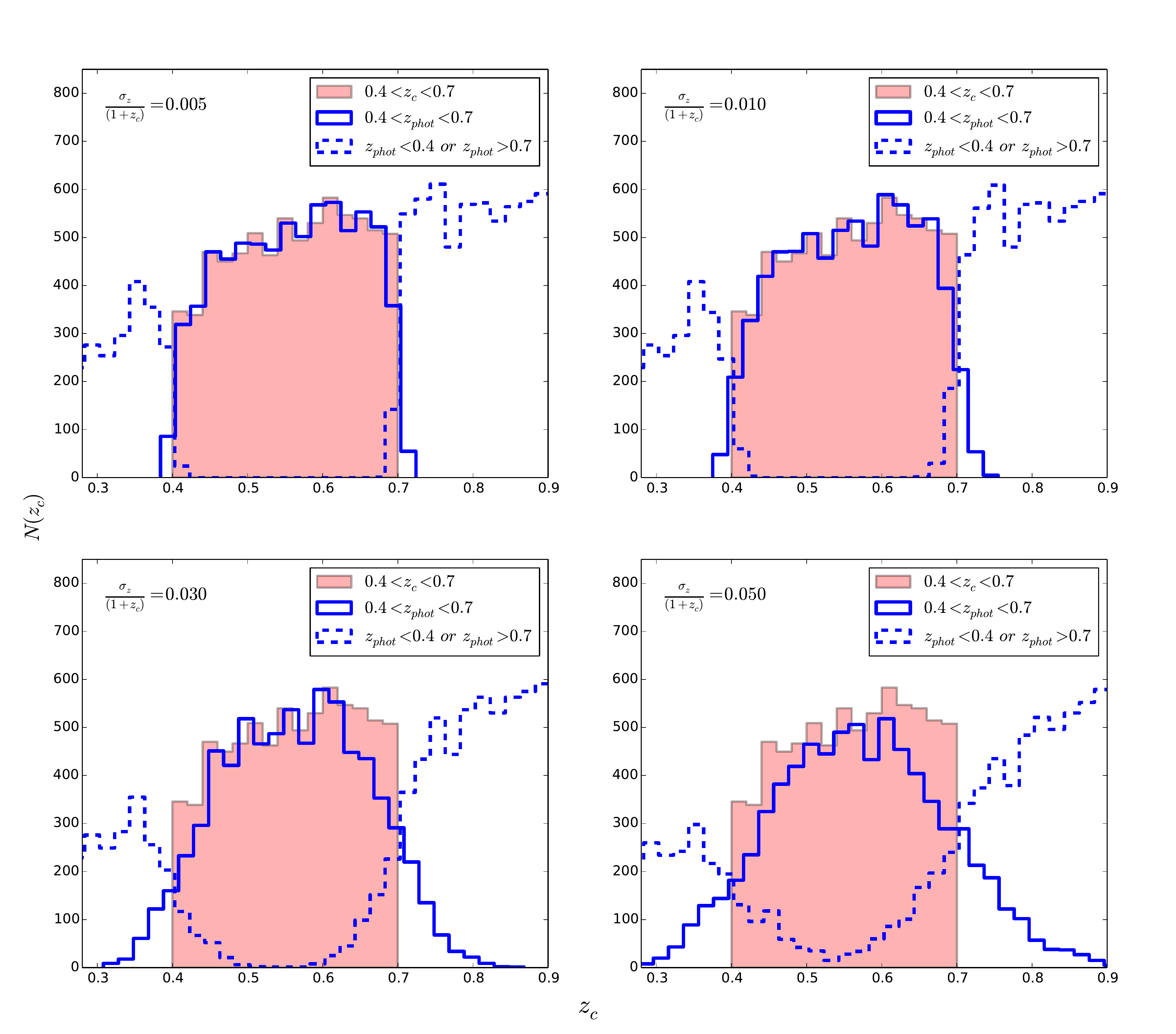}   
     \end{picture}
     \caption{Distribution of clusters selected in the top-hat cosmological redshift window compared 
     		 with the clusters selected in the top-hat photometric redshift window.
     		 Filled histograms correspond to
     		 distribution of clusters as a function of cosmological redshift when the top-hat
     		 selection is done using the cosmological redshift within the range $0.4<z<0.7$. Solid blue
     		 lines correspond to distribution of clusters as a function of cosmological redshift when the top-hat
     		 selection is done using the different photometric uncertainties we have used ($\sigma_{z}/(1+z_{c})$ = 
     		 0.005,0.010,0.030 and 0.050) with the range $0.4<z<0.7$ and the dashed blue lines correspond to 
     		 the distribution of clusters as a function of cosmological redshift when the top-hat
     		 selection is done using photometric redshifts outside the range $0.4<z<0.7$.}
     \label{fig:histogram_zcosmo_vs_zphot_0.4_z_0.7} 
\end{figure*}

\subsection{Recovering the real-space correlation function: the method}
\label{sec:deprojection_method}

\begin{figure*}
\centering
  \begin{tabular}{@{}cc@{}}
    \includegraphics[width=0.49\linewidth]{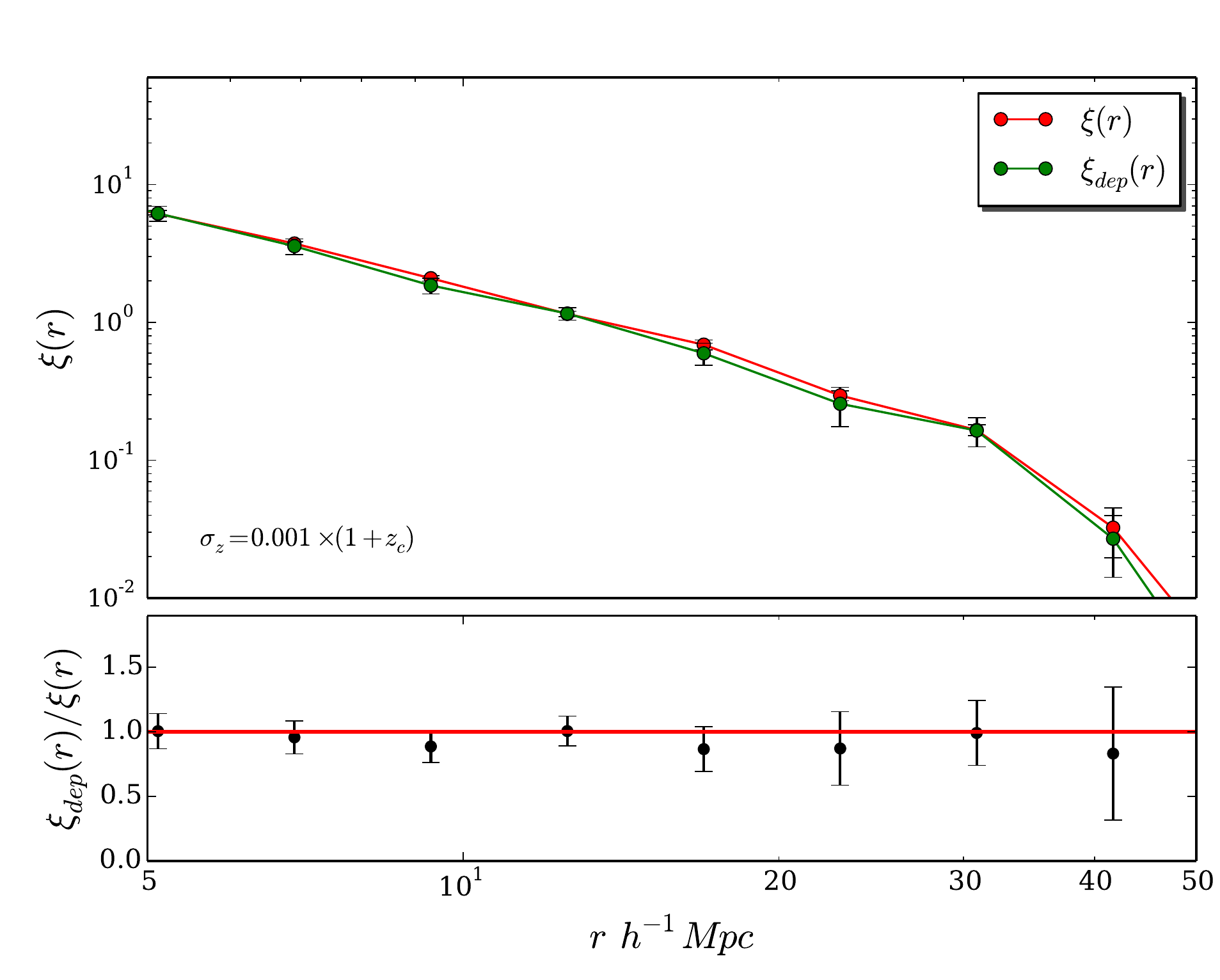} &
    \includegraphics[width=0.49\linewidth]{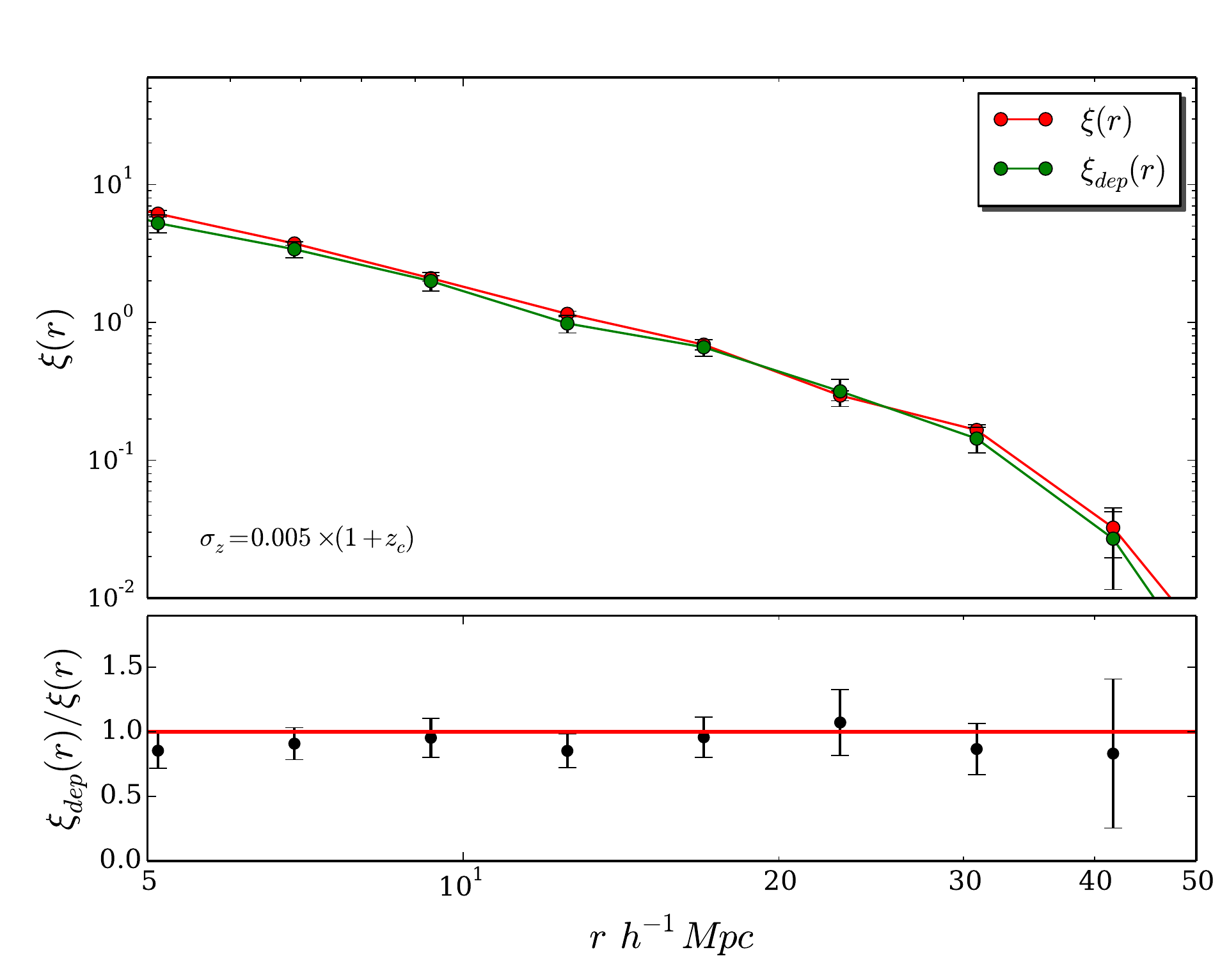}   \\
    \includegraphics[width=0.49\linewidth]{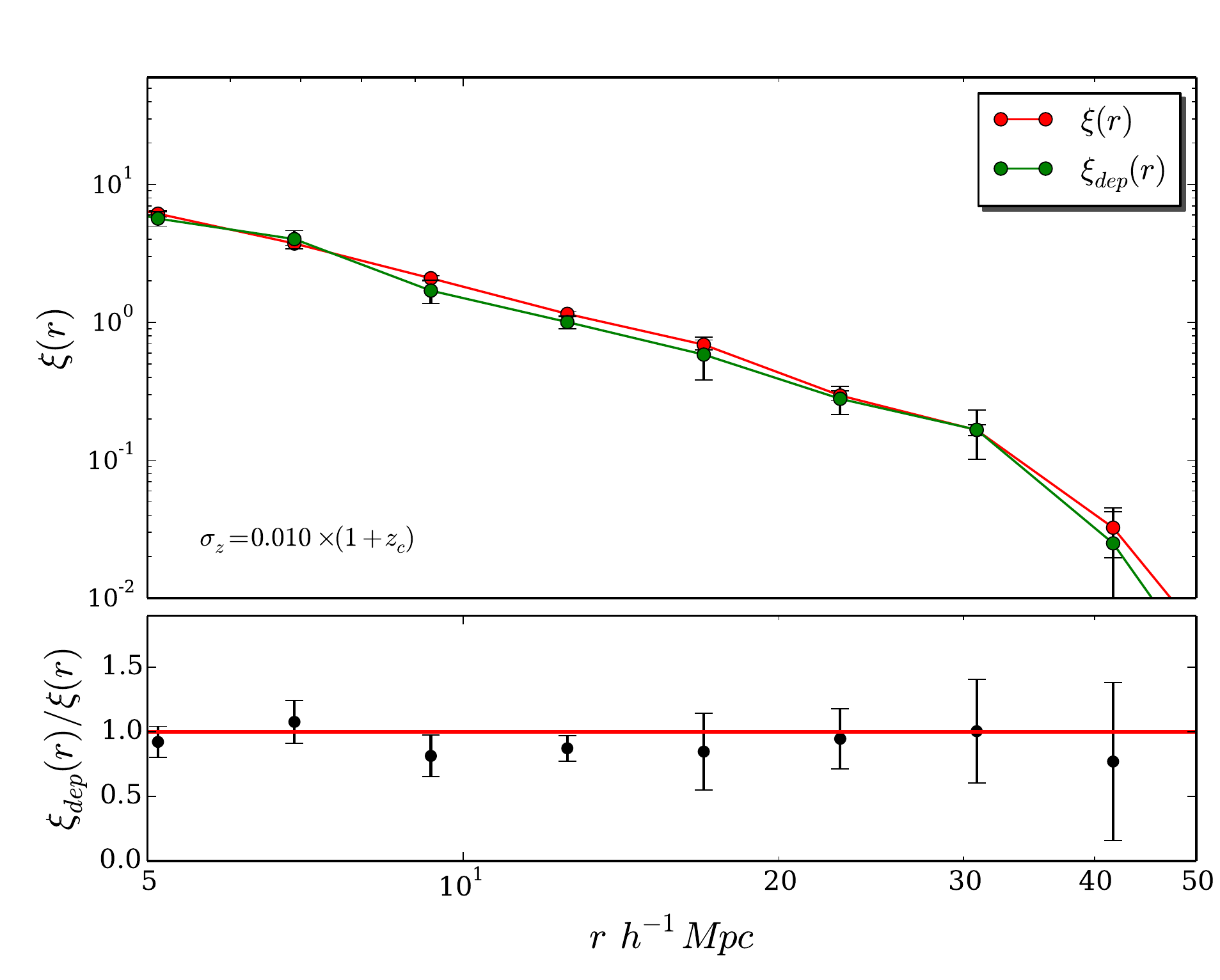} &
    \includegraphics[width=0.49\linewidth]{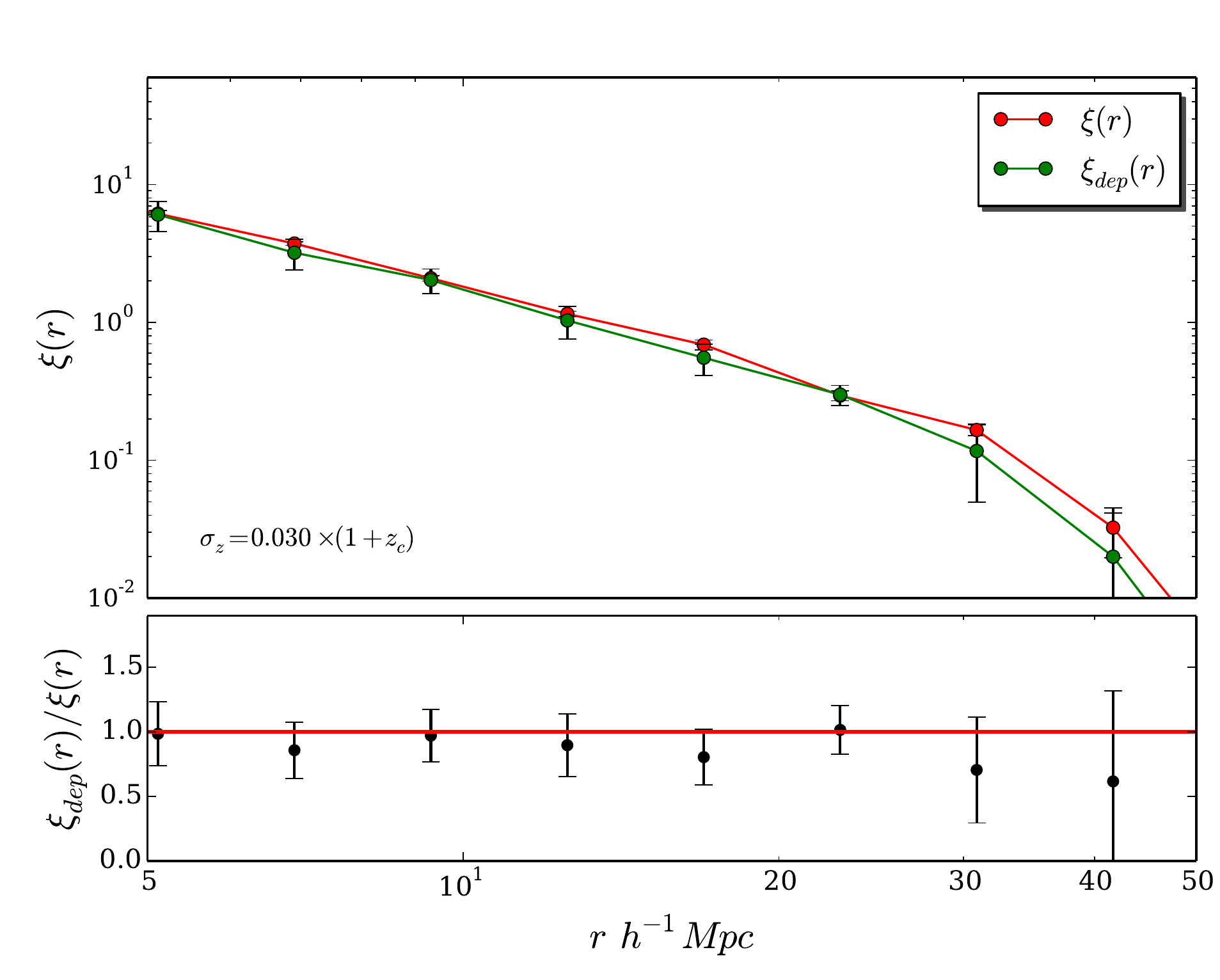}   \\
    \multicolumn{2}{c}{\includegraphics[width=0.49\linewidth]{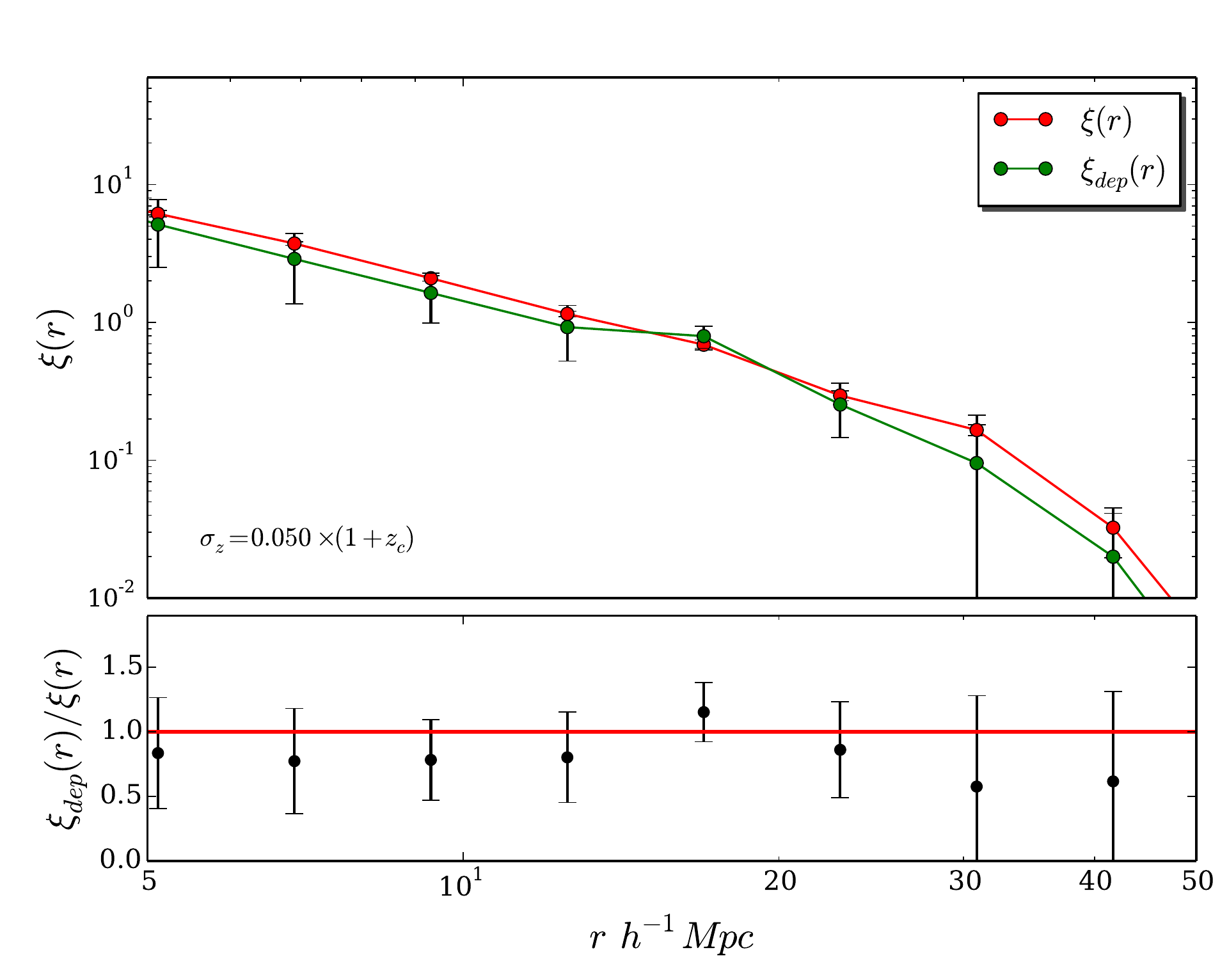}}
  \end{tabular}
  \caption{Recovered correlation function (green line) compared 
with the real-space correlation function (red line) for five mock 
photometric samples in the redshift range $0.4<z<0.7$, 
with increasing redshift uncertainty. 
Values of the best-fit parameters obtained are given in Table \ref{table:table_deltaxi} and 
the quality of the recovery for each sample is given in Table \ref{table:table_xidep}.}
\label{fig:results_plot}
\end{figure*}

\begin{figure*}
     \centering
     \setlength{\unitlength}{1cm}
     \begin{picture}(19,15.8)
     \includegraphics[width=\columnwidth]{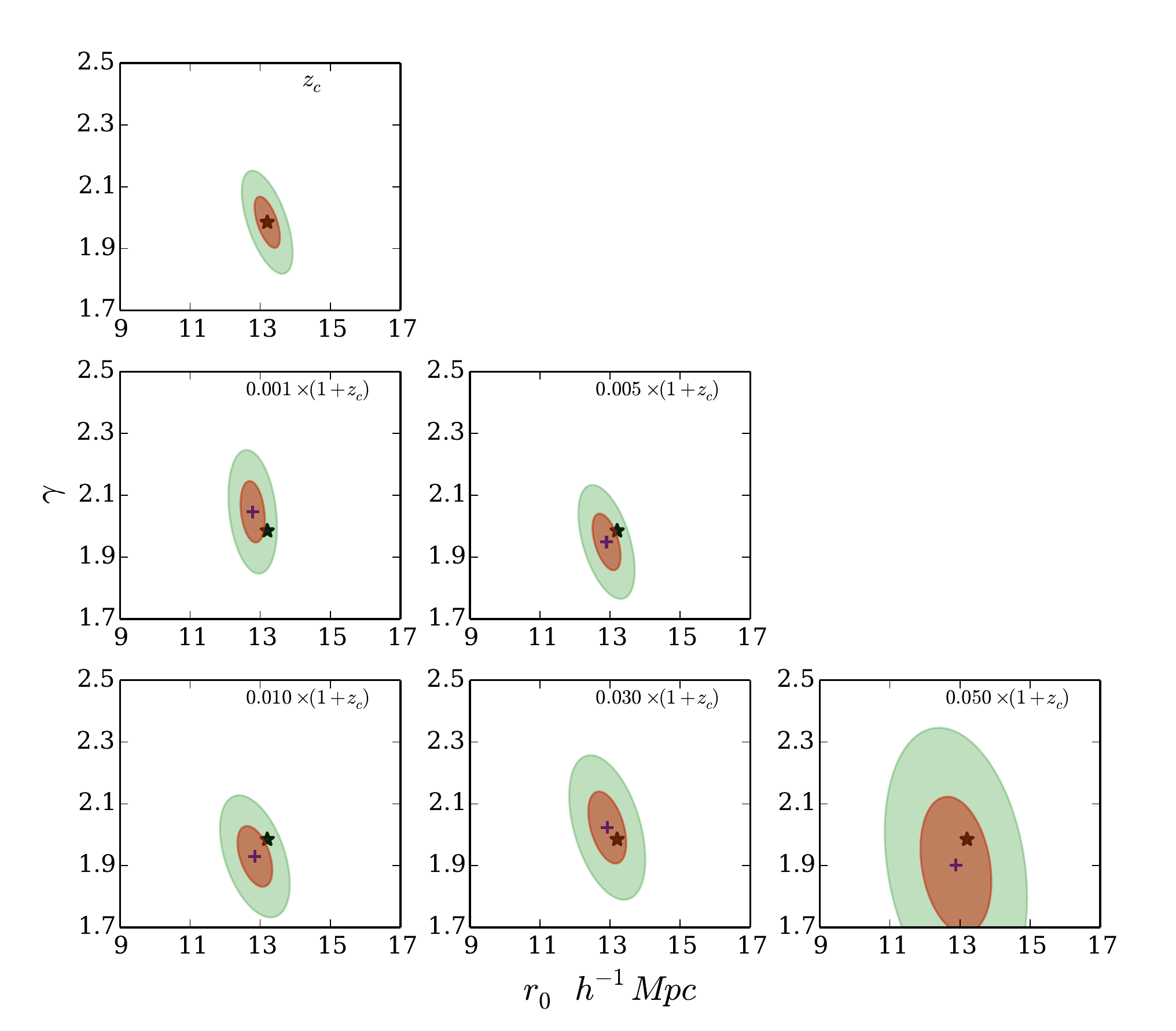}   
     \end{picture}
     \caption{1 $\sigma$  (shaded brown) and 3 $\sigma$ (shaded 
green) error ellipses for the parameters $r_{0}$ and $\gamma$.
Top panel: The original catalogue with cosmological redshifts
Central and bottom panels: Mock catalogues with increasing photometric 
redshift errors. The solid star represents the centre of the ellipse for the 
original catalogue, while the cross denotes the centres of the other 
ellipses.}
     \label{fig:error_ellipse_r0_vs_gamma}   
 \end{figure*}

\begin{table*}
\centering
\caption{Main parameters used for the analysis of the original catalogue
and the five mock photometric redshift catalogues: (1) the redshift uncertainty, 
(2) the maximum values of $\pi_{max}$ and (3) $r_{p(max)}$, (4) the values of $\Delta \xi$, (5) $\widehat{\Delta \xi}$. 
The range of scales $r$ used for the fit is fixed at 5-50 Mpc. 
} 

\label{table:table_deltaxi}
\begin{tabular}{*{6}{c}} 
\toprule
\multicolumn{1}{p{2cm}}{\centering Redshift uncertainty \\ $\left(\frac{\sigma_{z}}{1+z_c}\right)$}
& \multicolumn{1}{p{2cm}}{\centering $\pi_{max}$ \\ $(h^{-1}Mpc)$}
& \multicolumn{1}{p{2cm}}{\centering $r_{p(max)}$ \\ $(h^{-1}Mpc)$} & 
$\Delta \xi$ & $\widehat{\Delta \xi}$  \\ 
\midrule 
0.000 & 50 & 400 & 0.028 & 0.031 \\ [1ex]
0.001 & 60 & 400 & 0.042 & 0.052 \\ [1ex]
0.005 & 130 & 400 & 0.055 & 0.055  \\ [1ex]
0.010 & 300 & 400 & 0.065 & 0.063  \\ [1ex]
0.030 & 400 & 400 & 0.091 & 0.080  \\ [1ex]
0.050 & 550 & 400 & 0.148 & 0.109  \\ [1ex] \hline 
\end{tabular}
\end{table*}

\begin{table}
\centering
\caption{Best-fit parameters obtained for the real-space correlation 
function \xir{} of the original sample and the recovered deprojected 
correlation function $\xi_{dep}(r)$ for the mock photometric redshift samples. We quote the (1) redshift uncertainty, (2) the correlation length $r_{0}$, (3) slope $\gamma$. The mass cut used is $M_{halo}>5\times10^{13}\;h^{-1}\;M_{\odot}$
and the fit 
range is fixed at 5-50 Mpc. 
The fits have been performed both with fixed ($\gamma=2.0$) and free slope.} 
\label{table:table_xidep}
\begin{tabular}{*{3}{c}} 
\toprule

\multicolumn{1}{p{2cm}}{\centering Redshift uncertainty \\ $\left(\frac{\sigma_{z}}{1+z_c}\right)$}
& \multicolumn{1}{p{2cm}}{\centering $r_{0}$ \\ $(h^{-1}Mpc)$}
& \multicolumn{1}{p{2cm}}{\centering $\gamma$} \\ 
\midrule

\multirow{2}{*}{$z_c$} & 13.16$\pm$0.17 & 2.0 (\textit{fixed}) \\ [1ex] 
   		& 13.20$\pm$0.23 & 1.97$\pm$0.05  \\ [1ex] \hline \noalign{\vskip 0.1cm} 
   		   
\multirow{2}{*}{0.001} & 12.82$\pm$0.17 & 2.0 (\textit{fixed}) \\ [1ex] 
   		& 12.91$\pm$0.22 & 2.02$\pm$0.05  \\ [1ex] \hline \noalign{\vskip 0.1cm}
   		 
\multirow{2}{*}{0.005} & 12.52$\pm$0.22 & 2.0 (\textit{fixed}) \\ [1ex]
   		& 12.89$\pm$0.26 & 1.94$\pm$0.06  \\ [1ex] \hline \noalign{\vskip 0.1cm}
   		
\multirow{2}{*}{0.010} & 12.33$\pm$0.28 & 2.0 (\textit{fixed}) \\ [1ex]
   		& 12.84$\pm$0.63 & 1.93$\pm$0.08   \\ [1ex] \hline \noalign{\vskip 0.1cm}
   		
\multirow{2}{*}{0.030} & 12.29$\pm$0.30 & 2.0 (\textit{fixed}) \\ [1ex]
   		& 12.91$\pm$0.72 & 2.02$\pm$0.12   \\ [1ex] \hline \noalign{\vskip 0.1cm}
   		
\multirow{2}{*}{0.050} & 11.73$\pm$0.65 & 2.0 (\textit{fixed}) \\ [1ex]
   		& 12.88$\pm$0.76 & 1.90$\pm$0.14   \\ [1ex] \hline \noalign{\vskip 0.1cm} \\
\end{tabular}
\end{table}

In the following we will take into account separately 
the line of sight $\pi$ and the transverse $r_{p}$ components of the 
two--point correlation function.
Photometric redshifts affect only the line of sight component, 
introducing an anisotropy in the $\pi$-$r_{p}$ plane:
the redshift--space correlation function will have a lower
amplitude and steeper slope with respect to the real-space
correlation function \citep{pablo_arnalte}.

In order to recover the real-space correlation function of the photometric 
redshift mocks, we apply the deprojection method \citep{pablo_arnalte,Marulli_deprojection}.
The method is based on \citet{davis_peebles_1983} and \citet{saunders_1992}. 
Pairs are counted at different separations parallel ($\pi$) and perpendicular 
($r_{p}$) to the line of sight.

The comoving redshift space separation of the pair is defined as 
$\mathbf{s\equiv x_{2}-x_{1}}$ and the line of sight vector is 
$\mathbf{l = \frac{1}{2}(x_{1}+x_{2})}$
\citep{Fisher_1994}.  
The parallel and perpendicular distances to the 
pair are given by:

\begin{subequations}

\begin{equation}
\pi = \mathbf{\frac{s.l}{|l|}} ,
\end{equation}

\begin{equation}
r_{p} = \sqrt{\mathbf{|s|^{2}}-\pi^{2}} ,
\end{equation}

\end{subequations}
where $\bar{z} = \frac{1}{2}(z_{1}+z_{2})$. 
Counting pairs in both $(r_{p},\pi)$ dimensions will then provide  the 
anisotropic correlation function \xirppi.  
The projected correlation function can be derived from \xirppi{} by:

\begin{equation}\label{eqn:projected_correlation}
w_{p}(r_{p}) = \int_{-\infty}^{+\infty} \xi(r_{p},\pi)d\pi .
\end{equation}

The projected correlation function $w_{p}(r_{p})$ \citep{farrow} is related to the 
real-space correlation function \xir{} by Equation \ref{eqn:xir_wprp_relation}: 

\begin{equation}\label{eqn:xir_wprp_relation}
w_{p}(r_{p}) = 2\int_{r_{p}}^{\infty} rdr\xi(r)(r^{2}-r_{p}^{2})^{-1/2} ,
\end{equation}

which can be inverted to obtain the real-space correlation function:

\begin{equation}\label{eqn:recov_realspace}
\xi(r) = \frac{-1}{\pi} \int_{r}^{\infty} w'(r_{p})(r_{p}^{2}-r^{2})^{-1/2}dr_{p} .
\end{equation}

Theoretically, the upper limits of integration are infinite, but in practice
we need to choose finite values both in Equation 
\ref{eqn:projected_correlation} and Equation \ref{eqn:recov_realspace} 
which then become:

\begin{equation}\label{eqn:projected_correlation_pimax}
w_{p}(r_{p},\pi_{max}) = \int_{0}^{\pi_{max}} \xi(r_{p},\pi)d\pi ,
\end{equation}
and
\begin{equation}\label{eqn:recov_realspace_rpmax}
\xi(r) = \frac{-1}{\pi} \int_{r}^{r_{p_{max}}} w'(r_{p})(r_{p}^{2}-r^{2})^{-1/2}dr_{p} ,
\end{equation}

where $\pi_{max}$ and $r_{p_{max}}$  refer respectively to the maximum line of 
sight separation and the maximum transverse separation. 

Given that above a certain value of $\pi$, pairs are uncorrelated and \xirppi{} 
drops to zero, it is possible to find an optimal choice for $\pi_{max}$. 
This will be explained in detail in Section \ref{sec:limits}. 
We estimate the real-space correlation function
following the method of \citet{saunders_1992}. 
We use a step function to calculate $w_{p}(r_{p})$, where 
$w_{p}(r_{p})$ = $w_{p(i)}$ in the logarithmic interval centred on $r_{p(i)}$,
and we sum up in steps using the equation:

\begin{equation}
\begin{split}
\xi(r_{p(i)}) & = \frac{-1}{\pi} \sum_{j\geq i}
\frac{w_{p(j+1)}-w_{p(j)}}{r_{p(j+1)}-r_{p(j)}} \\
& ln \left( \frac{r_{p(j+1)} + \sqrt{r_{p(j+1)}^{2}-r_{p(i)}^{2}}}{r_{p(j)}+ 
\sqrt{r_{p(j)}^{2}-r_{p(i)}^{2}}} \right) .
\end{split}
\end{equation}

\begin{figure*}
\centering
  \begin{tabular}{@{}cc@{}}
    \includegraphics[width=0.49\linewidth]{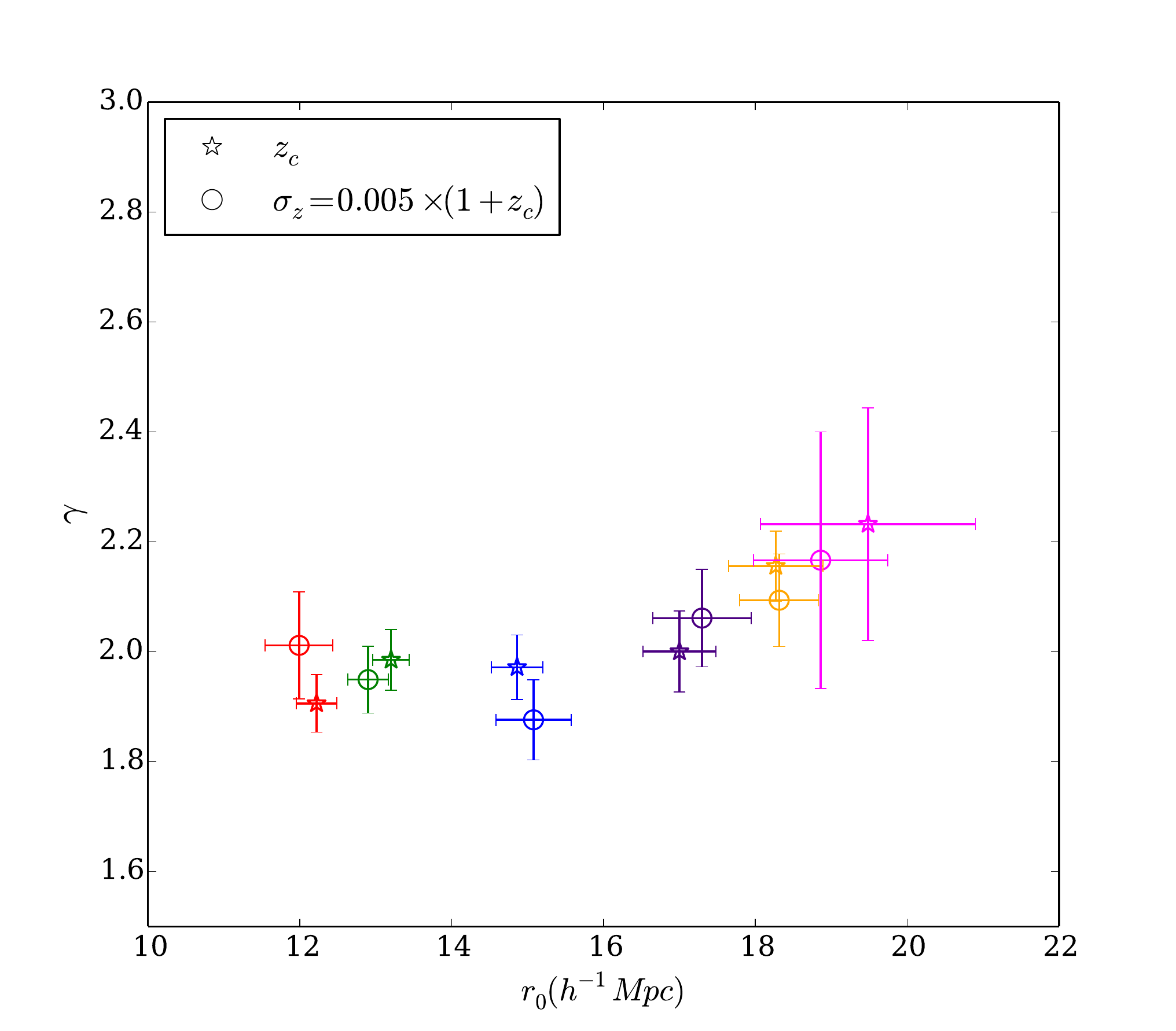} &
    \includegraphics[width=0.49\linewidth]{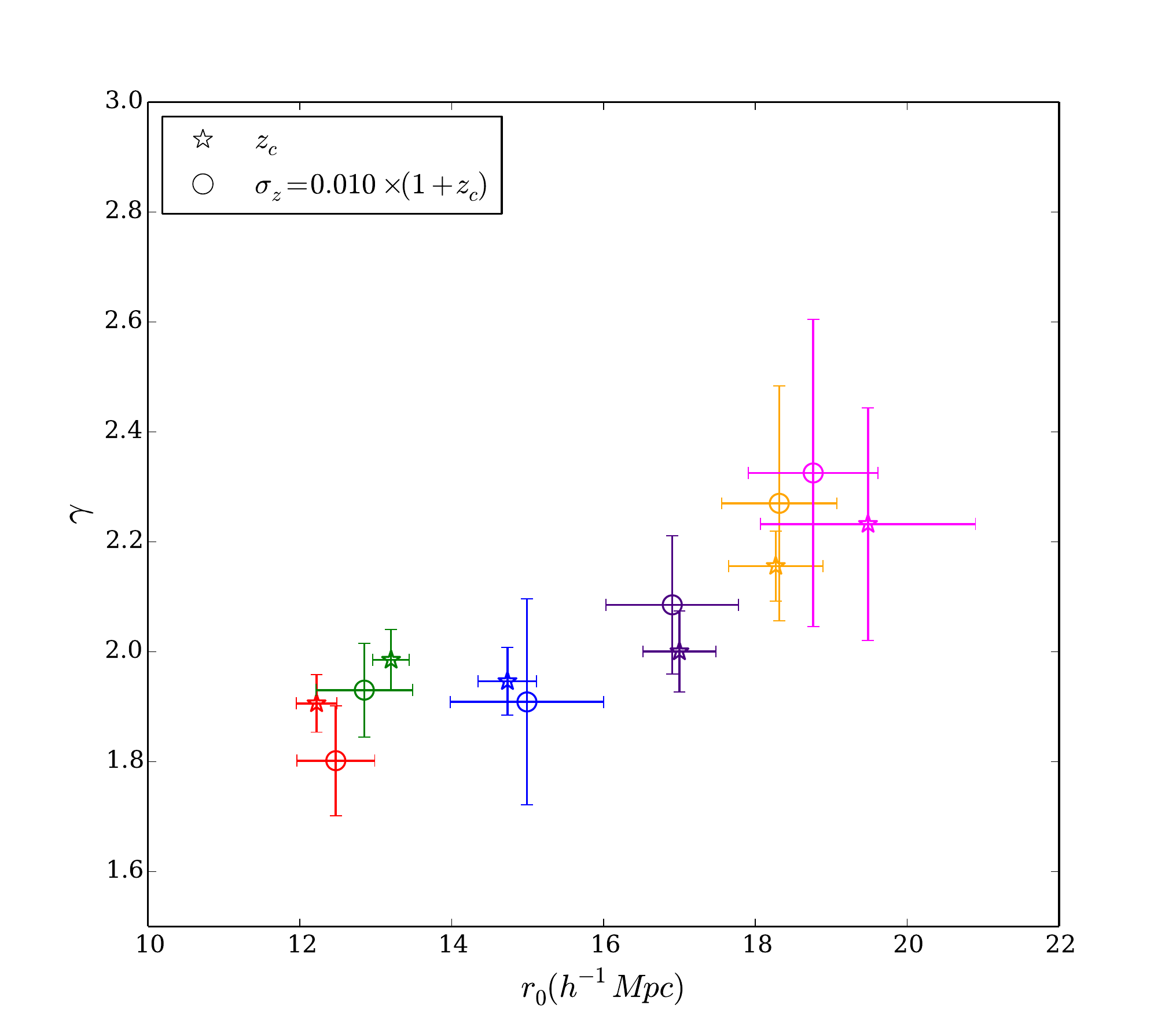}   \\
    \includegraphics[width=0.49\linewidth]{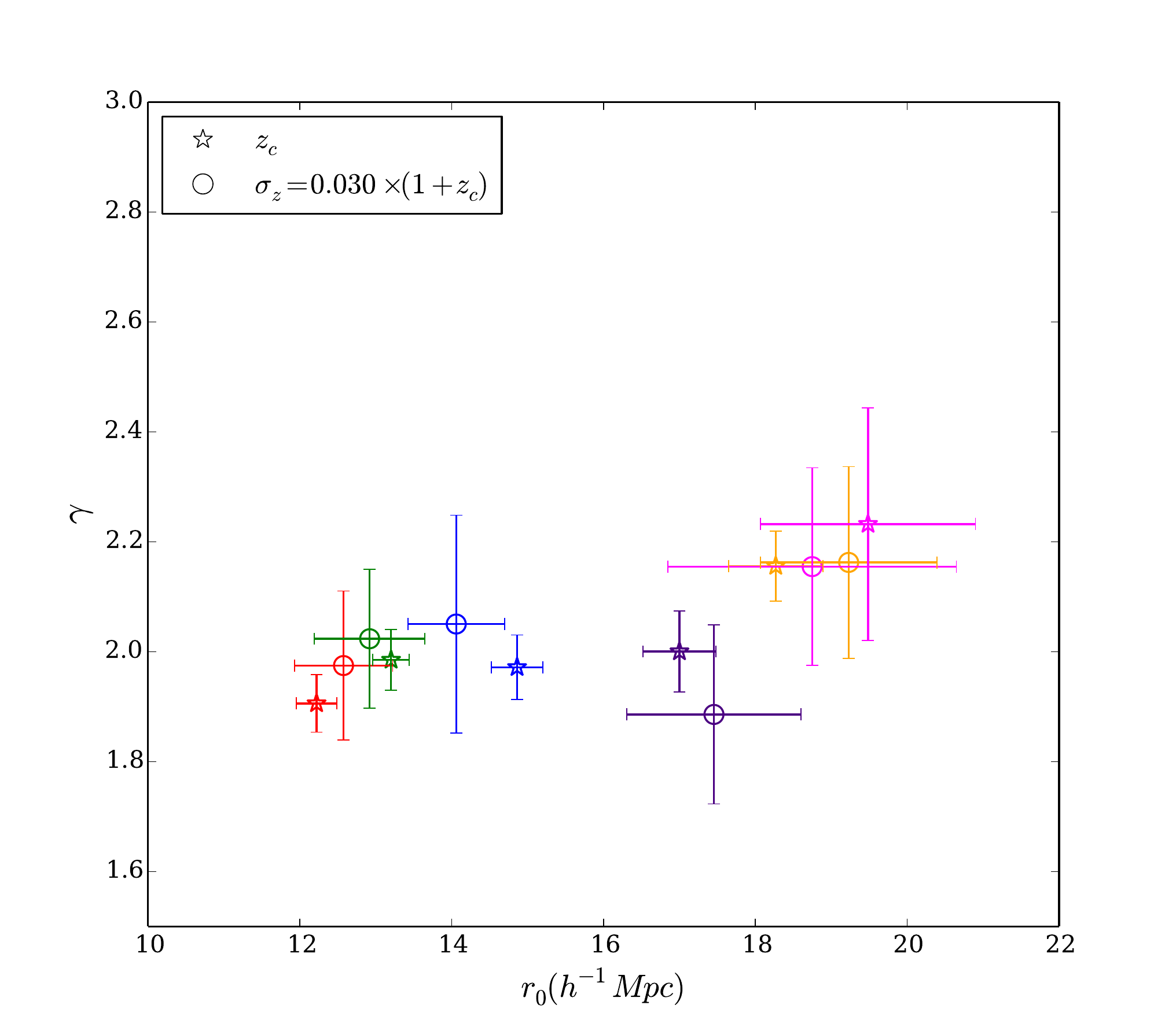} &
    \includegraphics[width=0.49\linewidth]{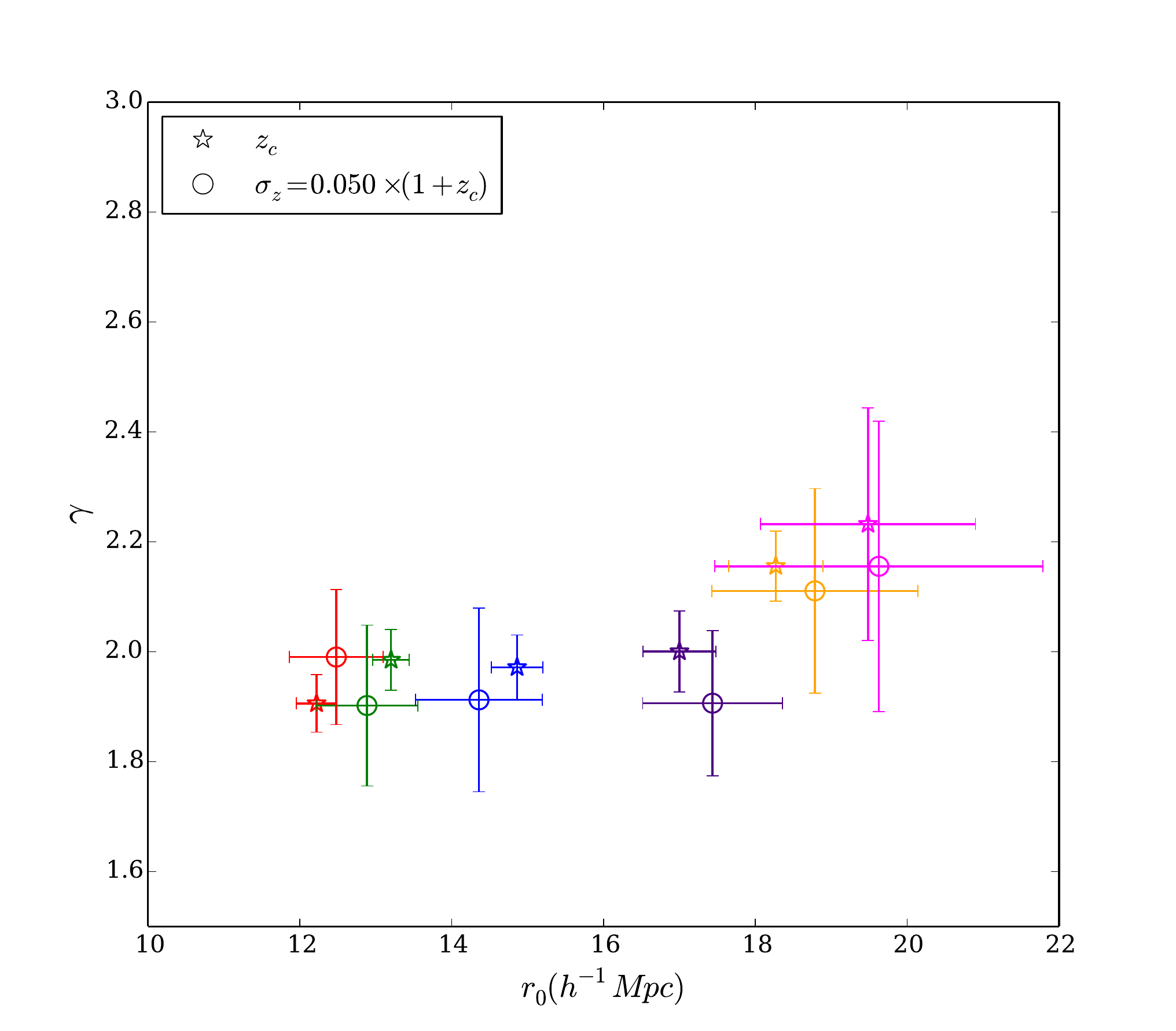}   \\
  \end{tabular}
  \caption{Evolution of $r_{0}$ and $\gamma$ with redshift for clusters with a mass cut 
$M_{halo}>5\times10^{13}\;h^{-1}\;M_{\odot}$ for samples with increasing redshift uncertainty 
($\sigma_{z}/(1+z_c)=0.005$, $0.010$, $0.030$ and $0.050$). Red $(0.1<z<0.4)$, 
Green $(0.4<z<0.7)$, Blue $(0.7<z<1.0)$, Indigo $(1.0<z<1.3)$, 
Gold $(1.3<z<1.6)$, Magenta $(1.6<z<2.1)$.}
\label{fig:z_vs_r0_zcosmo_vs_zphot}
\end{figure*}

Assuming that the correlation function follows a perfect power--law,
$w_{p}(r_{p})$ is given by the formula:

\begin{equation}\label{eqn:analytical_xir_wprp}
w_{p}(r_{p}) = r_{p}\left(\frac{r_{0}}{r_{p}}\right)^{\gamma} 
\frac{\Gamma(\frac{1}{2})\Gamma(\frac{\gamma -1}{2})}{\Gamma(\frac{\gamma}{2})} ,
\end{equation}
where $\Gamma$ is the Euler's gamma function. 

We compared the values of $r_{0}$ (with fixed slope) obtained 
from the fit of the recovered deprojected correlation function $\xi_{dep}(r)$ using Equation \ref{eqn:power-law-xir}. 
We also compared the values from the fit of $w_{p}(r_{p})$ with the same fixed slope using Equation 
\ref{eqn:analytical_xir_wprp} and found it to be similar to what we obtain for 
the recovered deprojected correlation function $\xi_{dep}(r)$.


\subsection{Application to a cluster mock catalogue}\label{sec:results}


\subsubsection{Photo-z catalogue selection}

As a first test, we applied the formalism described in the previous 
section to a mock cluster sample within the fixed redshift slice $0.4<z<0.7$.

For each cluster, we assigned a photometric redshift $z_{phot}$ following the probability 
$P(z_{phot}|z_c) =  G(z_c,\sigma_{z})$ where $\sigma_{z} = \sigma_{(z=0)} \times (1+z_{c})$.
As mentioned in \citet{Crocce_2011}, doing the selection in a top-hat photometric redshift window 
and in a top-hat cosmological redshift window with the same boundaries is not equivalent. 
Figure \ref{fig:histogram_zcosmo_vs_zphot_0.4_z_0.7} compares the distribution in cosmological 
redshift of the clusters selected in the top-hat cosmological redshift window $0.4<z_{c}<0.7$ 
(given by the filled histogram), the clusters selected by the top-hat photometric redshift window 
$0.4<z_{phot}<0.7$ (given by the solid blue line) and the clusters for which the photometric 
redshifts are outside the slice limits [0.4,0.7] (given by the dashed blue line) for four of 
our photometric samples. The distribution in cosmological redshift $N(z_{c})$ of the objects 
selected by the top-hat $z_{phot}$ window is broader than that selected by the top-hat $z_c$ window, 
and this effect increases with increasing $\sigma_z$. When performing the selection in $z_{phot}$ 
window rather than in $z_c$ window, a fraction of clusters with $z_c$ 
outside these slice limits but with $z_{phot}$ within the slice limits [0.4,0.7] are included, 
resulting then as contaminants. The distribution of clusters with $z_{phot}$ outside the window 
[0.4,0.7] is also shown as a dashed blue line. It shows that  a fraction of clusters with $z_{phot}$ 
outside [0.4,0.7]  have $z_c$  within the slice limits [0.4,0.7]. These objects are then lost by 
the top-hat photometric redshift selection.

The fraction of contaminating and missing clusters depends on the photometric redshift uncertainty 
and also on the $N(z)$ distribution. 
We calculate the fraction of common objects between the top-hat $z_{phot}$ and $z_c$ selections 
for the different $\sigma_{z}$ and redshift windows considered. It varies from 99\% to 70\% for 
samples with $\sigma_{z}/(1+z_{c})=0.001$ (at $z \approx 0.1 $) to $\sigma_{z}/(1+z_{c})=0.050$ (at $z \approx 1.3$) 
respectively. 
Only the samples with $\sigma_{z}/(1+z_{c})=0.050$ and above a redshift of $z>0.7$ 
have less than 80\% objects in common, 
as we know that the photo-z error scales as $\sigma_{z} = \sigma_{(z=0)} \times (1+z_{c})$. In our case there are 
four samples that fall into this category (fourth panel of 
Table \ref{table:contaminants_table}).
For all the other samples 
we chose, the average fraction of common clusters is more than 80\% and so by choosing a direct cut in 
photo-z space, we expect that the final clustering is not affected by a huge margin. To calculate the effect 
of $N(z)$ on contaminated and missing clusters, we calculate both the mean 
and median redshift for the photometric redshift samples we have. It can be seen from 
Table \ref{table:r0_vs_gamma_zcosmo_vs_zphot} that both the mean and the median redshift do not 
vary much when compared to the mean and median redshift of the cosmological redshift sample.  
The percentage of contaminants for each redshift slice and given photometric uncertainty along 
with the $N_{clusters}$ in $z_{c}$ and $z_{phot}$ window and the number of common clusters is 
mentioned in Table \ref{table:contaminants_table}.


\subsubsection{Selecting the integration limits}\label{sec:limits}

The \xirppi is calculated on a grid  with logarithmically 
spaced bins both in $r_{p}$ and $\pi$. The maximum value of
$r_{p}$ depends on the survey dimension in the transverse plane. 
In the redshift range $0.4<z<0.7$, the 
maximum separation across the line of sight direction in our light-cone 
is $\approx 500\;h^{-1}$Mpc. 
For the upper limit of integration in Equation \ref{eqn:recov_realspace_rpmax} 
we fixed a value $r_{p(max)} = 400\;h^{-1}$Mpc, corresponding to 80 \% of the
maximum transversal separation. 
For higher redshift samples we are aware that the maximum separation across the line of sight increases, 
but we find that the value of 400 $h^{-1}$Mpc includes almost all correlated pairs without adding any noise.

In the case of clusters where we have low statistics as compared to galaxy 
catalogues, the choice of the bin width must be taken into account, if not the 
Poisson noise will dominate.
A convergence test is performed for choosing the number and the width 
of bins in  $r_{p}$ and $\pi_{max}$.

Since higher photometric errors produce larger redshift space distortions, 
a different value of $\pi_{max}$ has to be fixed for each photometric redshift
mock. We determine its value in the following way.
We recover the real-space correlation function with the method 
described in Section \ref{sec:deprojection_method}, using increasing values of 
$\pi_{max}$. Initially the amplitude of $\xi_{dep}(r)$ is underestimated because 
many correlated pairs are not taken into account; it 
increases when increasing $\pi_{max}$ up to a maximum value, beyond which it 
starts to fluctuate and noise starts to dominate.  
Applying this test to each mock, we select the
$\pi_{max}$ value corresponding to the maximum recovered amplitude.
 
\begin{figure}[!ht]
\centering
\includegraphics[width=\linewidth]{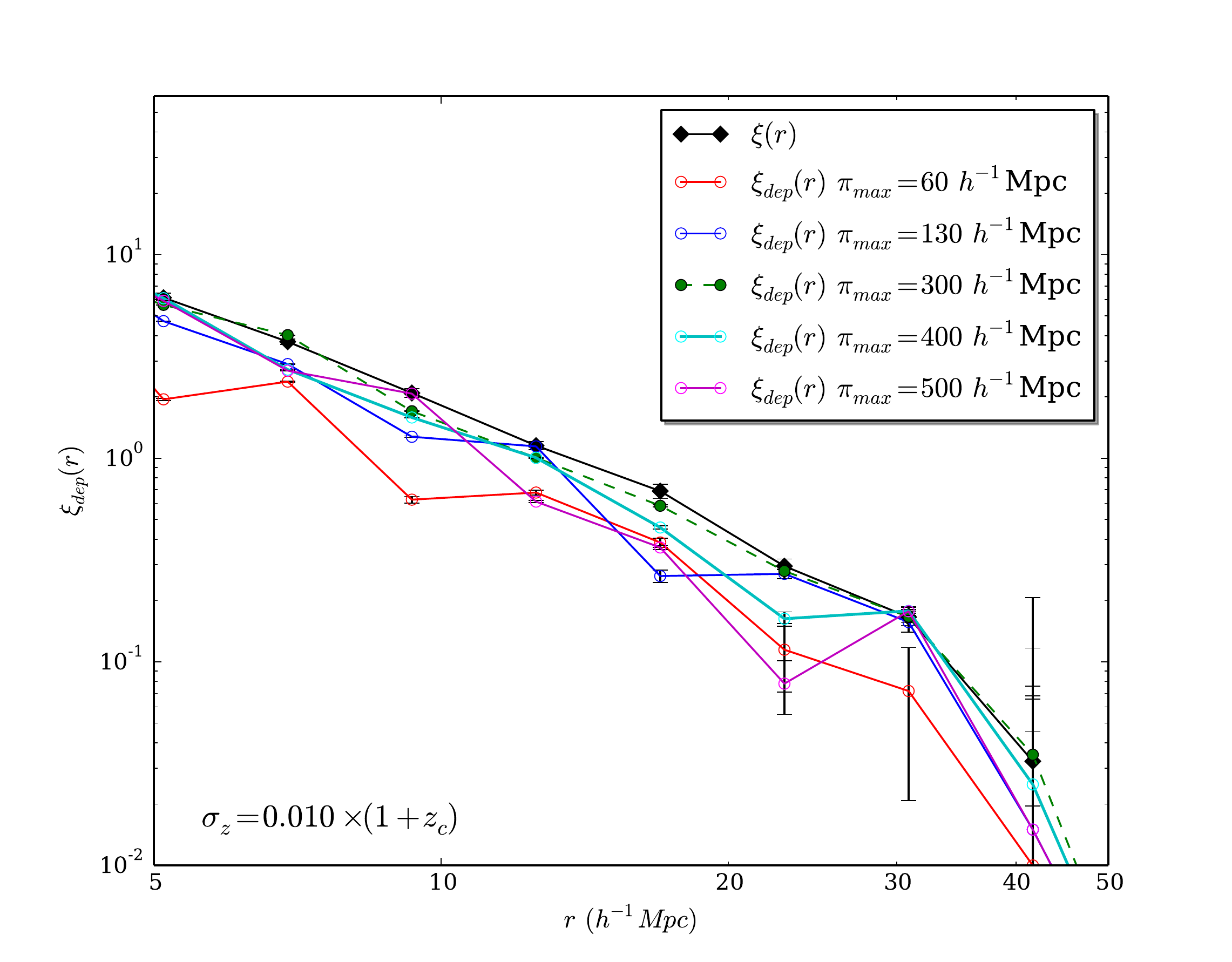}
\caption{Recovered correlation function with different values of $\pi_{max}$ 
(as colour coded in the figure) for the sample with $\sigma_{z} = 0.010\times(1+z_{c})$ in the redshift 
range $0.4<z<0.7$. The black line joining the 
diamonds in both the plots is the real-space correlation function calculated for the cosmological redshift 
sample (same as the red line in Figure \ref{fig:results_plot}).
Poisson error bars are plotted just for convenience.}
\label{fig:convergence_zphot1}
\end{figure}

We show an example of the $\pi_{max}$ test for the
photometric sample with $\sigma_{z} = 0.010\times(1+z_{c})$.
Figure \ref{fig:convergence_zphot1} shows that the amplitude of
the correlation function increases with increasing $\pi_{max}$, but only up to a certain value, 
which we call the maximum recovered amplitude. 
It can be seen that integrating the function above this value of $\pi_{max}$ only results 
in noise.

It is clear from our tests on simulations (see Figure \ref{fig:convergence_zphot1}) that there is an
optimal $\pi_{max}$ value; integrating beyond that limit increases the noise.
In future work on observed cluster samples, using the data
themselves, we can examine the value of the observed correlation amplitude as
a function of $\pi_{max}$, choosing the $\pi_{max}$ value providing the maximum
correlation amplitude.

We have checked that by applying this method to the original light-cone with 
cosmological redshifts we correctly recover its real-space correlation 
function. The values of  $\pi_{max}$ used for our reference sample ($0.4<z<0.7$) are 
given in Table \ref{table:table_deltaxi}.


\subsubsection{The quality of the recovery}

In Figure \ref{fig:results_plot} we compare
$\xi_{dep}(r)$ of our five mocks with the real-space correlation function 
\xir{}.
It is clear that  $\xi_{dep}(r)$ reproduces quite well \xir{}, but shows
increasing fluctuations with increasing $\sigma_{z}$. 
The ratio  $\xi_{dep}(r) / \xi(r)$ is slightly smaller than one but within 
$1 \sigma$ at all scales for all the mocks 
up to $\sigma_z/(1+z_{c})  \: = \: 0.05$.
 
The quality of the recovery is determined using  $\Delta \xi$, 
an ``average normalised residual''  defined by \citet{pablo_arnalte} as:

\begin{equation}
\Delta\xi = \frac{1}{N}\sum_{i}\bigg|\frac{\xi_{dep}(r_{i})-
\xi(r_{i})}{\xi(r_{i})}\bigg| ,
\end{equation} 
where $r_{i}$ refers to the values in the $i^{th}$ bin considered and
$\xi(r_{i})$ is the real-space correlation function. 

In the case of real data, where $z_{c}$ is not available,
one can still calculate the quality of the recovery using the covariance matrix and is defined as:

\begin{equation}
 \widehat{\Delta \xi} = \frac{1}{N}\sum_{i} \frac{\sqrt{C_{ii}}}{|\xi_{dep}(r_{i})|} ,
\end{equation}  
wherein we use the covariance matrix that we have obtained using the jackknife resampling method mentioned in 
Equation \ref{eqn:jackknife}.
The values of $\Delta \xi$ and $\widehat{\Delta \xi}$ 
estimated in the range 5-50 $h^{-1}$Mpc, are listed in Table 
\ref{table:table_deltaxi}.

One can see  from Table \ref{table:table_deltaxi} that for the lowest photometric error considered,
$\sigma_z/(1+z_{c}) \: = \: 0.001$, the real-space correlation function 
is recovered within 5\%. For $\sigma_z/(1+z_{c}) \: = \: 0.005$ and 
$\sigma_z/(1+z_{c}) \: = \: 0.010$ it is recovered within 7\%, within 9\% for $\sigma_z/(1+z_{c}) \: = \: 0.030$,
and finally within 15\% for $\sigma_z/(1+z_{c}) \: = \: 0.05$.

The best-fit parameters of the deprojected correlation functions are shown in Table \ref{table:table_xidep}. 
The fitting is performed with both a 
free and fixed slope $\gamma=2.0$. 
The correlation length obtained for our five mock photometric samples is 
consistent within $\sim 1 \sigma$ with the real-space correlation length
$r_{0} = 13.20\pm0.23\;h^{-1}$Mpc and $r_{0} = 13.16\pm0.17\;h^{-1}$Mpc  
obtained for the $z_c$ sample for \xir{} (free slope) and \xir{} (fixed slope) respectively. 
The best-fit $r_{0}$ obtained for this particular sample ($0.4<z_{c}<0.7$) seems to have a value that is 
always lower, regardless of the photometric uncertainty, when compared to the $r_{0}$ obtained for the true 
$z_{c}$ sample. This is just a coincidence and is not always the case, as can be seen for other 
samples with different redshift limits. 
When the slope is set free, direct comparison of $r_{0}$ between the samples cannot be made and so  
in Figure  \ref{fig:error_ellipse_r0_vs_gamma} we plot the three sigma error 
ellipses around the best-fit values of
$r_{0}$ and $\gamma$ for all the mocks. 
As expected, the errors on both $r_{0}$ and $\gamma$ increase with the 
photometric error, but are always within $\sim 1 \sigma$ with respect to the real
space values.

We also applied the deprojection method for higher photometric redshift errors 
to test how far  the method could be applied. It was found that from  $\sigma_z/(1+z_{c}) = 0.1$, 
the error on the recovery is very large and the recovered correlation function becomes biased.


\subsubsection{Recovering the redshift evolution of the correlation function from sub-samples selected using photometric redshifts}

We checked how accurately we can follow the redshift 
evolution of the cluster real-space correlation function 
when using photometric redshifts and the deprojection method to retrieve the
real-space correlation function.
We have previously shown this for the light-cone with cosmological redshifts in Figure \ref{fig:r0_vs_gamma_zcosmo}. 
 
For this purpose, we analysed four mocks with redshift uncertainties of 
$\sigma_z/(1+z_{c}) \: = \: 0.005, 0.010, 0.030$, and $0.050$ respectively, in five redshift slices, 
from $0.1 < z <0.4$ to $1.6 < z < 2.1$ with the same mass cut $M_{halo}>5\times10^{13}\;h^{-1}\;M_{\odot}$ 
as done in Section \ref{sec:Redshift_evolution}. 
The results are shown in Figure \ref{fig:z_vs_r0_zcosmo_vs_zphot} and the 
values of the best-fit parameters for all the four photometric samples are given 
in Table \ref{table:r0_vs_gamma_zcosmo_vs_zphot} in the Appendix section
along with the number of clusters ($N_{clusters}$) and the mean and median redshift for each sample.

\begin{figure*}
	\centering
     \setlength{\unitlength}{1cm}
     \begin{picture}(24, 16)
     \includegraphics[scale=0.49]{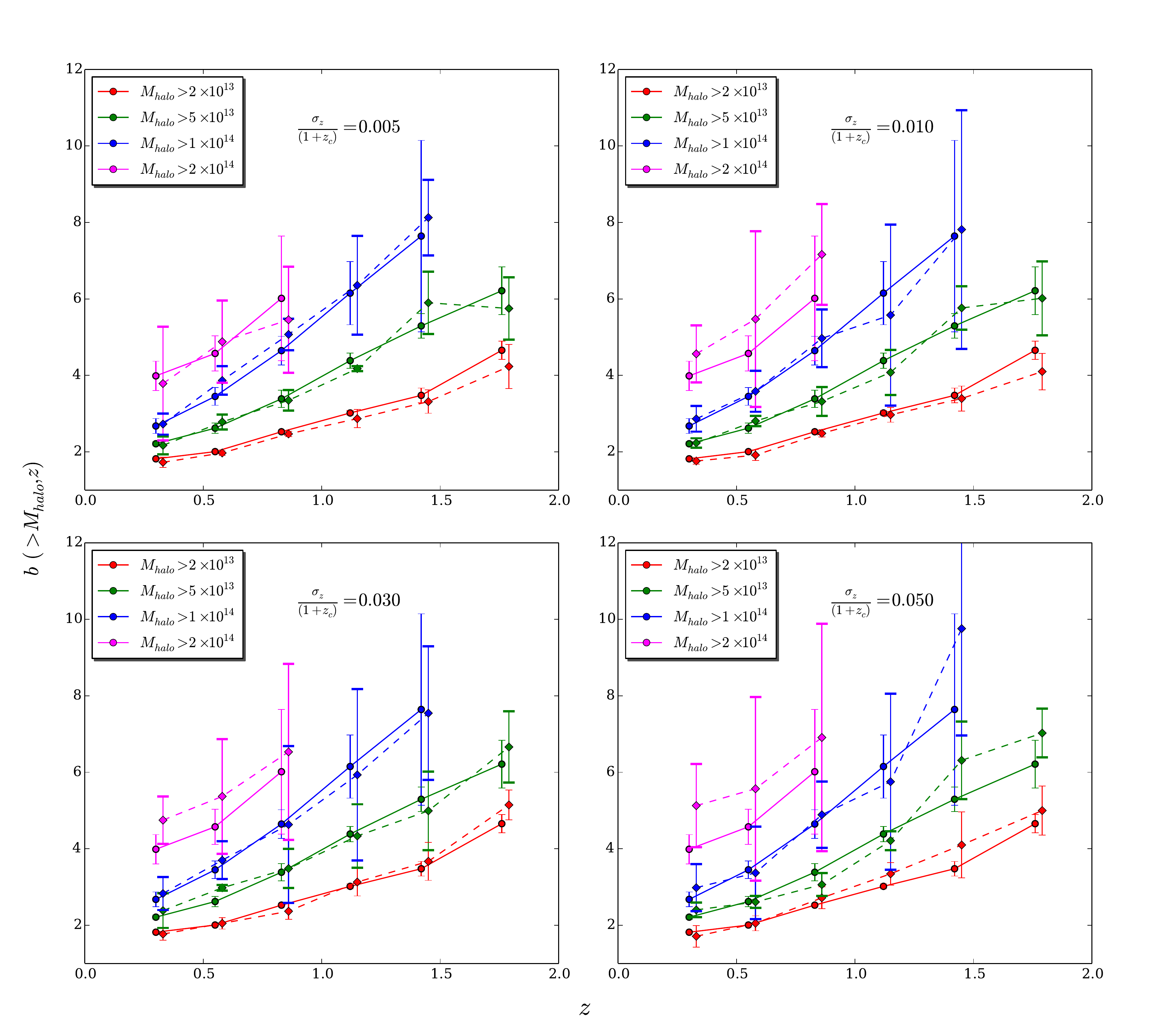}   
     \end{picture}
     \caption{Evolution of bias with redshift and mass (with units $h^{-1}\;M_{\odot}$) 
for the $z_c$ sample (solid lines) compared with the photometric samples (dashed lines) 
with redshift uncertainties of $\sigma_z/(1+z_{c})=0.005,0.010,0.030$ and $\sigma_z/(1+z_{c}) = 0.050$. }
     \label{fig:z_vs_bias_zcosmo_vs_zphot} 
\end{figure*}

Figure \ref{fig:z_vs_r0_zcosmo_vs_zphot} shows the evolution of the best-fit parameters $
r_{0}$ and $\gamma$ for the different redshift slices. 
The four panels correspond to the different photometric redshift errors tested. It can be compared to 
Figure \ref{fig:r0_vs_gamma_zcosmo} which shows the values of $r_{0}$ and $\gamma$ estimated for the same 
redshift slices but using cosmological redshifts. The fits are performed in the range within 
which $\xi(r)$ can be described using a power-law.
As in Figure \ref{fig:r0_vs_gamma_zcosmo}, $r_0$ and $\gamma$ are shown to increase with 
redshift but the errors on their 
estimates become larger as the photometric redshift error increases. As a result, the 
error bars on the $r_0$ estimates for consecutive redshift slices tested tend to superimpose when 
considering large  values of $\sigma_z$. The increase of $r_0$ with redshift remains detectable, 
but a larger binning in redshift is needed to detect this effect  significantly when working 
with large $\sigma_z$.  
We note that the parameters estimated from the deprojected correlation function are within 1$\sigma$ 
from the ones estimated directly in real-space, and that remains true even for high redshifts and for high 
values of the photometric errors. The large error bars for the last two redshift slices 
($1.3<z<1.6$ and $1.6<z<2.1$) are both due to the small number
of clusters at high redshift (see the histogram shown in Figure 
\ref{fig:hist_random_vs_cluster_dist}) and the scaling of the photometric error 
$\sigma_{z} = \sigma_{(z=0)}\times (1+z_c)$. However, we can see that the low number of clusters 
makes the correlation function hard to measure even using cosmological redshifts.  
From our tests we can conclude that the correlation function can be recovered from photometric 
redshift surveys using the deprojection 
method up to a redshift of $z \approx 2.0$  within $10\%$ percent 
with a photometric redshift error of $\sigma_z/(1+z_{c}) = 0.030$. In this sense, the recovery 
performed with this method can be considered as successful.
Even in the last redshift slice chosen ($1.6<z<2.1$), the correlation function can be recovered within $1\sigma$
for all the four photometric
redshift uncertainties tested.  It can be numerically visualised
in the last panel of Table \ref{table:r0_vs_gamma_zcosmo_vs_zphot}. This point is of particular 
importance as the $1.5<z<2.0$ redshift range has been shown to be very discriminant for 
constraining cosmological parameters with clusters \citep{Sartoris_2016}.

We also estimated the bias as defined in Section \ref{sec:bias} for 
$\sigma_z/(1+z_{c}) = 0.005,0.010,0.030$, and $0.050$. Our values are given 
in Tables \ref{table:bias_values_zphot_1_and_2} and 
\ref{table:bias_values_zphot_3_and_4} in the Appendix section along with the number of clusters
($N_{clusters}$) in each sample.
The results can be 
seen in Figure \ref{fig:z_vs_bias_zcosmo_vs_zphot}. 
The bias values obtained for the photo-z samples are consistent
with the values obtained for the reference sample and are within $1\sigma$ error bars. 
For the first two mass cut samples ($M_{halo}>2\times 10^{13}\;h^{-1}\;M_{\odot}$ and 
$5\times 10^{13}\;h^{-1}\;M_{\odot}$), the calculated bias from 
the photometric samples are within 1$\sigma$ even up to a median redshift of $z \approx 1.8$.
Up to a mass cut of 
$M_{halo}>1\times 10^{14}\;h^{-1}\;M_{\odot}$ one can see that the redshift evolution of the bias 
can be traced very well 
(even up to redshifts of $z \approx 1.5$).

However we notice that for the highest mass cut sample ($M_{halo}>2\times 10^{14}\;h^{-1}\;M_{\odot}$) 
chosen, only the bias values
obtained for the photometric sample with $\sigma_z/(1+z_{c}) = 0.005$ seem to be similar to that obtained by the 
reference sample. The remaining three photometric samples depict a much higher bias (even though
they fall within $1\sigma$) when compared to the reference sample. One reason for this behaviour and
also for the large error bars for this mass cut sample is 
due to the smaller abundance of clusters at this mass threshold cut as it can be seen from 
Table \ref{table:bias_values_zphot_1_and_2} and \ref{table:bias_values_zphot_3_and_4}. We also believe that
it could be due to the percentage of contaminants that are present in this mass cut sample for three 
different photometric uncertainties. We have calculated the contaminants 
for this mass cut sample and they seem to be higher at certain 
redshifts when compared to the contaminants at the same redshifts found for the low mass cut samples.

However, up to a mass cut of $M_{halo}>5\times 10^{13}\;h^{-1}\;M_{\odot}$, the evolution in redshift 
and mass of the bias is 
clearly distinguished, that too up to the highest redshift tested ($z \approx 2.1$).


\section{Discussion and conclusions}\label{sec:conclusion}

In future, most of the cluster detections in large galaxy surveys will be 
based on photometric catalogues.
Therefore the main aim of this work was to 
apply a method for recovering the spatial two--point correlation function 
$\xi(r)$ of clusters using only photometric redshifts and assess 
its performance.

In order to estimate the real-space correlation function $\xi(r)$, 
we applied a method originally developed to correct for peculiar velocity 
distortions, first estimating
the projected correlation function \wprp{}, then applying Equation 
\ref{eqn:recov_realspace_rpmax}  to deproject it.

For our analysis we used the 500 deg$^{2}$ light-cone of \citet{alex_merson}.
Mock photometric redshifts were generated from the 
cosmological redshifts assuming a Gaussian error 
$\sigma_z = \sigma_{(z=0)}\times (1+z_c)$ (as described in Section \ref{sec:mock_photo_samples}). 
This represents a first approximation, sufficient for the scope of the present 
work; a more realistic approach will have to 
include real photo-z distributions and catastrophic failures.

Here are our main results.

\begin{enumerate}

\item We directly estimate the cluster
correlation function in real-space (i.e. using cosmological redshifts) 
for sub-samples of the light-cone in different redshift intervals
and with different mass thresholds
(see Section \ref{sec:Redshift_evolution} and \ref{sec:clustering_with_mass}). 
As expected, we find an increasing clustering
strength with both redshift and mass threshold. 
At a fixed mass threshold, the 
correlation amplitude increases with redshift, while at a fixed redshift the
correlation amplitude increases with mass threshold.
The increase of the correlation amplitude with redshift is larger
for more massive haloes: for example, for
$M_{halo}>2\times 10^{13}\;h^{-1}\;M_{\odot}$, $r_{0} = 9.89\pm0.20$ at 
$z = 0.25$ and $r_{0} = 12.41\pm0.42$ at $z = 1.45$; for 
$M_{halo}>1\times 10^{14}\;h^{-1}\;M_{\odot}$, $r_{0} = 14.60\pm0.35$ at 
$z = 0.25$ and $r_{0} = 26.09\pm4.10$ at $z = 1.45$. 

\item We fit the relation between the clustering length $r_{0}$ 
and the mean intercluster distance $d$ in the redshift interval 
$0.1 \le z \le  2.1$
up to $z \approx 2.0$, finding
$r_0 = 1.77\pm0.08(d)^{0.58\pm0.01} \; h^{-1}$Mpc, which is  consistent with the 
relation $r_0 = 1.70(d)^{0.60} \; h^{-1}$Mpc obtained by \citet{Younger} for
the local redshift range $0 \le z \le 0.3$. 

\item We estimate the bias parameter directly in real-space 
(using cosmological redshifts), with different mass thresholds. 
Analogously to the correlation
amplitude, the bias increases with redshift, and the increase is larger
for more massive clusters. 
Our results are consistent with \citet{estrada} and with
the theoretical prediction of \citet{Tinker_2010}.

\item We finally apply the deprojection method to recover the 
real-space correlation function $\xi(r)$ of different sub-samples using
photometric redshifts. We  recover $\xi(r)$ 
within $\sim$ 7\% on scales $5<r<50\;h^{-1}$Mpc
with a photometric error of
$\sigma_z/(1+z_{c}) = 0.010$ and within $\sim$ 9\%  for samples with 
$\sigma_z/(1+z_{c}) = 0.030$; the best--fit parameters of the recovered real--space 
correlation function, as well as the bias, are within $1\sigma$ of the 
corresponding values for the direct estimate in real-space, up to $z \sim 2$.
\end{enumerate}

Our results are promising in view of future surveys such as Euclid and 
LSST that will provide state-of-the-art  photometric redshifts over an 
unprecedented range of redshift scales. 
This work represents the first step towards a more complete analysis taking into account 
the different observational problems
to be faced when determining cluster clustering from real data.  We are planning to extend 
this study to the  use of more realistic photo-z errors. 
We are also planning to apply the deprojection method to cluster catalogues produced
by  cluster detection algorithms. This implies taking into account the selection function of the 
cluster catalogue, and investigating the impact of purity and completeness 
on clustering. Another important issue to be faced is that mass in general is not available for 
cluster catalogues derived from data so that a proxy of mass such as richness has to be used. 
The fact that the scatter of the relation between mass and richness 
introduces another uncertainty  has to be taken into account when using 
clusters for constraining the cosmological parameters
(see e.g. 
\citet{Berlind_2003_n200, Kravtsov_2004_n200, Zheng_2005_n200, Rozo_2009_n200, 
Rykoff_2012_n200}).
Another important constraint for cosmological parameters is given by
the BAO feature in the two-point correlation function 
\citep{Veropalumbo_1,Veropalumbo_2}. As we have pointed out in 
Section \ref{sec:data}, the size of the light-cone we used 
(500 deg$^{2}$) is not large enough to detect the BAO feature. 
It will be interesting to extend this analysis to forthcoming all-sky simulations to 
test if the BAO feature can be detected using photometric redshifts, and if so with what accuracy, 
in next generation surveys.


\begin{acknowledgements}
We thank Dr. Alex Merson (previously at University College London) and Carlton Baugh (Durham University) 
for fruitful discussions on the simulation used in this work.
We thank Dr. Pablo Arnalte-Mur (Universitat de Val\`{e}ncia) for a detailed discussion on the deprojection 
method used and Gianluca Castignani for giving us insights into the subjects of error analysis 
and other statistical methods used in this research. 
We would also like to thank the anonymous referee for helpful comments. 

The authors acknowledge the Euclid Consortium, the European Space Agency and the support of the agencies and 
institutes that have supported the development of Euclid.  A detailed complete list is available on the 
Euclid web site (http://www.euclid-ec.org).

We thank in particular the Agenzia Spaziale Italiana, the Centre National d'Etudes Spatiales, the Deutches Zentrum ' 
fur Luft- and Raumfahrt, the Danish Space Research Institute, the Funda{\c c}{\~a}o para a Cienca e a Tecnologia, 
the Ministerio de Economia y Competitividad, the National Aeronautics and Space Administration, the 
Netherlandse Onderzoekschool Voor Astronomie, the Norvegian Space Center, the Romanian Space Agency, the 
United Kingdom Space Agency and the University of Helsinki.

Srivatsan Sridhar is supported by the Erasmus Mundus Joint Doctorate Program by 
Grant Number 2013-1471 from the agency EACEA of the European Commission.
Christophe Benoist, Sophie Maurogordato and Srivatsan Sridhar acknowledge financial support from CNES/INSU131425 grant. 
This research made use of TOPCAT and STIL: Starlink Table/VOTable Processing Software developed by \citet{topcat} 
and also the Code for Anisotropies in the Microwave Background (CAMB) \citep{CAMB_1,CAMB_2}. 

The Millennium Simulation databases \citep{Lemson_2006} used in this paper and the web application 
providing online access to them were constructed as part of the activities of the German Astrophysical 
Virtual Observatory.

Srivatsan Sridhar would also like to thank Sridhar Krishnan, Revathy Sridhar and Madhumitha Srivatsan 
for their support and encouragement during this work.
\end{acknowledgements}

\bibliographystyle{aa}
\bibliography{recovery_paper}

\begin{thebibliography}{74}
\expandafter\ifx\csname natexlab\endcsname\relax\def\natexlab#1{#1}\fi

\bibitem[{{Angulo} {et~al.}(2005){Angulo}, {Baugh}, {Frenk}, {Bower},
  {Jenkins}, \& {Morris}}]{Angulo_2005}
{Angulo}, R.~E., {Baugh}, C.~M., {Frenk}, C.~S., {et~al.} 2005, \mnras, 362,
  L25

\bibitem[{{Arnalte-Mur} {et~al.}(2009){Arnalte-Mur}, {Fern{\'a}ndez-Soto},
  {Mart{\'{\i}}nez}, {Saar}, {Hein{\"a}m{\"a}ki}, \&
  {Suhhonenko}}]{pablo_arnalte}
{Arnalte-Mur}, P., {Fern{\'a}ndez-Soto}, A., {Mart{\'{\i}}nez}, V.~J., {et~al.}
  2009, \mnras, 394, 1631

\bibitem[{{Ascaso} {et~al.}(2015){Ascaso}, {Mei}, \& {Ben{\'{\i}}tez}}]{Ascaso}
{Ascaso}, B., {Mei}, S., \& {Ben{\'{\i}}tez}, N. 2015, \mnras, 453, 2515

\bibitem[{{Bahcall} {et~al.}(2003){Bahcall}, {Dong}, {Hao}, {Bode}, {Annis},
  {Gunn}, \& {Schneider}}]{Bahcall_2003}
{Bahcall}, N.~A., {Dong}, F., {Hao}, L., {et~al.} 2003, \apj, 599, 814

\bibitem[{{Bahcall} {et~al.}(2004){Bahcall}, {Hao}, {Bode}, \&
  {Dong}}]{Bahcall_2004}
{Bahcall}, N.~A., {Hao}, L., {Bode}, P., \& {Dong}, F. 2004, \apj, 603, 1

\bibitem[{{Bahcall} \& {Soneira}(1983)}]{Bahcall_soneira_d}
{Bahcall}, N.~A. \& {Soneira}, R.~M. 1983, \apj, 270, 20

\bibitem[{{Bahcall} \& {West}(1992)}]{Bahcall_west}
{Bahcall}, N.~A. \& {West}, M.~J. 1992, \apj, 392, 419

\bibitem[{{Berlind} {et~al.}(2003){Berlind}, {Weinberg}, {Benson}, {Baugh},
  {Cole}, {Dav{\'e}}, {Frenk}, {Jenkins}, {Katz}, \&
  {Lacey}}]{Berlind_2003_n200}
{Berlind}, A.~A., {Weinberg}, D.~H., {Benson}, A.~J., {et~al.} 2003, \apj, 593,
  1

\bibitem[{{Bode} \& {Ostriker}(2003)}]{Bode_Ostriker}
{Bode}, P. \& {Ostriker}, J.~P. 2003, \apjs, 145, 1

\bibitem[{{Borgani} {et~al.}(1999){Borgani}, {Plionis}, \&
  {Kolokotronis}}]{Borgani}
{Borgani}, S., {Plionis}, M., \& {Kolokotronis}, V. 1999, \mnras, 305, 866

\bibitem[{{Bower} {et~al.}(2006){Bower}, {Benson}, {Malbon}, {Helly}, {Frenk},
  {Baugh}, {Cole}, \& {Lacey}}]{Bower_2006}
{Bower}, R.~G., {Benson}, A.~J., {Malbon}, R., {et~al.} 2006, \mnras, 370, 645

\bibitem[{{Colberg} {et~al.}(2000){Colberg}, {White}, {Yoshida}, {MacFarland},
  {Jenkins}, {Frenk}, {Pearce}, {Evrard}, {Couchman}, {Efstathiou}, {Peacock},
  {Thomas}, \& {Virgo Consortium}}]{Colberg_2000}
{Colberg}, J.~M., {White}, S.~D.~M., {Yoshida}, N., {et~al.} 2000, \mnras, 319,
  209

\bibitem[{{Crocce} {et~al.}(2011){Crocce}, {Cabr{\'e}}, \&
  {Gazta{\~n}aga}}]{Crocce_2011}
{Crocce}, M., {Cabr{\'e}}, A., \& {Gazta{\~n}aga}, E. 2011, \mnras, 414, 329

\bibitem[{{Croft} {et~al.}(1997){Croft}, {Dalton}, {Efstathiou}, {Sutherland},
  \& {Maddox}}]{Croft_d}
{Croft}, R.~A.~C., {Dalton}, G.~B., {Efstathiou}, G., {Sutherland}, W.~J., \&
  {Maddox}, S.~J. 1997, \mnras, 291, 305

\bibitem[{{Davis} {et~al.}(1985){Davis}, {Efstathiou}, {Frenk}, \&
  {White}}]{Davis_from_Merson}
{Davis}, M., {Efstathiou}, G., {Frenk}, C.~S., \& {White}, S.~D.~M. 1985, \apj,
  292, 371

\bibitem[{{Davis} \& {Peebles}(1983)}]{davis_peebles_1983}
{Davis}, M. \& {Peebles}, P.~J.~E. 1983, \apj, 267, 465

\bibitem[{{Eisenstein} {et~al.}(2011){Eisenstein}, {Weinberg}, {Agol},
  {Aihara}, {Allende Prieto}, {Anderson}, {Arns}, {Aubourg}, {Bailey},
  {Balbinot}, \& et~al.}]{SDSS}
{Eisenstein}, D.~J., {Weinberg}, D.~H., {Agol}, E., {et~al.} 2011, \aj, 142, 72

\bibitem[{{Estrada} {et~al.}(2009){Estrada}, {Sefusatti}, \&
  {Frieman}}]{estrada}
{Estrada}, J., {Sefusatti}, E., \& {Frieman}, J.~A. 2009, \apj, 692, 265

\bibitem[{{Farrow} {et~al.}(2015){Farrow}, {Cole}, {Norberg}, {Metcalfe},
  {Baldry}, {Bland-Hawthorn}, {Brown}, {Hopkins}, {Lacey}, {Liske}, {Loveday},
  {Palamara}, {Robotham}, \& {Sridhar}}]{farrow}
{Farrow}, D.~J., {Cole}, S., {Norberg}, P., {et~al.} 2015, \mnras, 454, 2120

\bibitem[{{Fisher} {et~al.}(1994){Fisher}, {Davis}, {Strauss}, {Yahil}, \&
  {Huchra}}]{Fisher_1994}
{Fisher}, K.~B., {Davis}, M., {Strauss}, M.~A., {Yahil}, A., \& {Huchra}, J.~P.
  1994, \mnras, 267, 927

\bibitem[{{Font} {et~al.}(2008){Font}, {Bower}, {McCarthy}, {Benson}, {Frenk},
  {Helly}, {Lacey}, {Baugh}, \& {Cole}}]{Font_2008}
{Font}, A.~S., {Bower}, R.~G., {McCarthy}, I.~G., {et~al.} 2008, \mnras, 389,
  1619

\bibitem[{{Garilli} {et~al.}(2014){Garilli}, {Guzzo}, {Scodeggio},
  {Bolzonella}, {Abbas}, {Adami}, {Arnouts}, {Bel}, {Bottini}, {Branchini},
  {Cappi}, {Coupon}, {Cucciati}, {Davidzon}, {De Lucia}, {de la Torre},
  {Franzetti}, {Fritz}, {Fumana}, {Granett}, {Ilbert}, {Iovino}, {Krywult}, {Le
  Brun}, {Le F{\`e}vre}, {Maccagni}, {Ma{\l}ek}, {Marulli}, {McCracken},
  {Paioro}, {Polletta}, {Pollo}, {Schlagenhaufer}, {Tasca}, {Tojeiro},
  {Vergani}, {Zamorani}, {Zanichelli}, {Burden}, {Di Porto}, {Marchetti},
  {Marinoni}, {Mellier}, {Moscardini}, {Nichol}, {Peacock}, {Percival},
  {Phleps}, \& {Wolk}}]{VIPERS_1}
{Garilli}, B., {Guzzo}, L., {Scodeggio}, M., {et~al.} 2014, \aap, 562, A23

\bibitem[{{Gonzalez-Perez} {et~al.}(2014){Gonzalez-Perez}, {Lacey}, {Baugh},
  {Lagos}, {Helly}, {Campbell}, \& {Mitchell}}]{Gonzalez_2014}
{Gonzalez-Perez}, V., {Lacey}, C.~G., {Baugh}, C.~M., {et~al.} 2014, \mnras,
  439, 264

\bibitem[{{Governato} {et~al.}(1999){Governato}, {Babul}, {Quinn}, {Tozzi},
  {Baugh}, {Katz}, \& {Lake}}]{Governato_d}
{Governato}, F., {Babul}, A., {Quinn}, T., {et~al.} 1999, \mnras, 307, 949

\bibitem[{{Guo} {et~al.}(2013){Guo}, {White}, {Angulo}, {Henriques}, {Lemson},
  {Boylan-Kolchin}, {Thomas}, \& {Short}}]{Guo_2013}
{Guo}, Q., {White}, S., {Angulo}, R.~E., {et~al.} 2013, \mnras, 428, 1351

\bibitem[{{Guzzo} {et~al.}(2014){Guzzo}, {Scodeggio}, {Garilli}, {Granett},
  {Fritz}, {Abbas}, {Adami}, {Arnouts}, {Bel}, {Bolzonella}, {Bottini},
  {Branchini}, {Cappi}, {Coupon}, {Cucciati}, {Davidzon}, {De Lucia}, {de la
  Torre}, {Franzetti}, {Fumana}, {Hudelot}, {Ilbert}, {Iovino}, {Krywult}, {Le
  Brun}, {Le F{\`e}vre}, {Maccagni}, {Ma{\l}ek}, {Marulli}, {McCracken},
  {Paioro}, {Peacock}, {Polletta}, {Pollo}, {Schlagenhaufer}, {Tasca},
  {Tojeiro}, {Vergani}, {Zamorani}, {Zanichelli}, {Burden}, {Di Porto},
  {Marchetti}, {Marinoni}, {Mellier}, {Moscardini}, {Nichol}, {Percival},
  {Phleps}, \& {Wolk}}]{VIPERS_2}
{Guzzo}, L., {Scodeggio}, M., {Garilli}, B., {et~al.} 2014, \aap, 566, A108

\bibitem[{{Hamilton}(1993)}]{hamilton}
{Hamilton}, A.~J.~S. 1993, \apj, 417, 19

\bibitem[{{Hong} {et~al.}(2012){Hong}, {Han}, {Wen}, {Sun}, \&
  {Zhan}}]{Hong_2012}
{Hong}, T., {Han}, J.~L., {Wen}, Z.~L., {Sun}, L., \& {Zhan}, H. 2012, \apj,
  749, 81

\bibitem[{{Hopkins} {et~al.}(2005){Hopkins}, {Bahcall}, \& {Bode}}]{Hopkins}
{Hopkins}, P.~F., {Bahcall}, N.~A., \& {Bode}, P. 2005, \apj, 618, 1

\bibitem[{Howlett {et~al.}(2012)Howlett, Lewis, Hall, \& Challinor}]{CAMB_2}
Howlett, C., Lewis, A., Hall, A., \& Challinor, A. 2012, JCAP, 1204, 027

\bibitem[{{Huchra} {et~al.}(1990){Huchra}, {Henry}, {Postman}, \&
  {Geller}}]{Huchra}
{Huchra}, J.~P., {Henry}, J.~P., {Postman}, M., \& {Geller}, J., M. 1990, \apj,
  365, 66

\bibitem[{{H{\"u}tsi}(2010)}]{Hutsi}
{H{\"u}tsi}, G. 2010, \mnras, 401, 2477

\bibitem[{{Ilbert} {et~al.}(2006){Ilbert}, {Arnouts}, {McCracken},
  {Bolzonella}, {Bertin}, {Le F{\`e}vre}, {Mellier}, {Zamorani}, {Pell{\`o}},
  {Iovino}, {Tresse}, {Le Brun}, {Bottini}, {Garilli}, {Maccagni}, {Picat},
  {Scaramella}, {Scodeggio}, {Vettolani}, {Zanichelli}, {Adami}, {Bardelli},
  {Cappi}, {Charlot}, {Ciliegi}, {Contini}, {Cucciati}, {Foucaud}, {Franzetti},
  {Gavignaud}, {Guzzo}, {Marano}, {Marinoni}, {Mazure}, {Meneux}, {Merighi},
  {Paltani}, {Pollo}, {Pozzetti}, {Radovich}, {Zucca}, {Bondi}, {Bongiorno},
  {Busarello}, {de La Torre}, {Gregorini}, {Lamareille}, {Mathez}, {Merluzzi},
  {Ripepi}, {Rizzo}, \& {Vergani}}]{Ilbert_2006}
{Ilbert}, O., {Arnouts}, S., {McCracken}, H.~J., {et~al.} 2006, \aap, 457, 841

\bibitem[{{Ilbert} {et~al.}(2009){Ilbert}, {Capak}, {Salvato}, {Aussel},
  {McCracken}, {Sanders}, {Scoville}, {Kartaltepe}, {Arnouts}, {Le Floc'h},
  {Mobasher}, {Taniguchi}, {Lamareille}, {Leauthaud}, {Sasaki}, {Thompson},
  {Zamojski}, {Zamorani}, {Bardelli}, {Bolzonella}, {Bongiorno}, {Brusa},
  {Caputi}, {Carollo}, {Contini}, {Cook}, {Coppa}, {Cucciati}, {de la Torre},
  {de Ravel}, {Franzetti}, {Garilli}, {Hasinger}, {Iovino}, {Kampczyk},
  {Kneib}, {Knobel}, {Kovac}, {Le Borgne}, {Le Brun}, {F{\`e}vre}, {Lilly},
  {Looper}, {Maier}, {Mainieri}, {Mellier}, {Mignoli}, {Murayama}, {Pell{\`o}},
  {Peng}, {P{\'e}rez-Montero}, {Renzini}, {Ricciardelli}, {Schiminovich},
  {Scodeggio}, {Shioya}, {Silverman}, {Surace}, {Tanaka}, {Tasca}, {Tresse},
  {Vergani}, \& {Zucca}}]{Ilbert_2009}
{Ilbert}, O., {Capak}, P., {Salvato}, M., {et~al.} 2009, \apj, 690, 1236

\bibitem[{{Ivezic} {et~al.}(2008){Ivezic}, {Tyson}, {Abel}, {Acosta},
  {Allsman}, {AlSayyad}, {Anderson}, {Andrew}, {Angel}, {Angeli}, {Ansari},
  {Antilogus}, {Arndt}, {Astier}, {Aubourg}, {Axelrod}, {Bard}, {Barr},
  {Barrau}, {Bartlett}, {Bauman}, {Beaumont}, {Becker}, {Becla}, {Beldica},
  {Bellavia}, {Blanc}, {Blandford}, {Bloom}, {Bogart}, {Borne}, {Bosch},
  {Boutigny}, {Brandt}, {Brown}, {Bullock}, {Burchat}, {Burke}, {Cagnoli},
  {Calabrese}, {Chandrasekharan}, {Chesley}, {Cheu}, {Chiang}, {Claver},
  {Connolly}, {Cook}, {Cooray}, {Covey}, {Cribbs}, {Cui}, {Cutri}, {Daubard},
  {Daues}, {Delgado}, {Digel}, {Doherty}, {Dubois}, {Dubois-Felsmann},
  {Durech}, {Eracleous}, {Ferguson}, {Frank}, {Freemon}, {Gangler}, {Gawiser},
  {Geary}, {Gee}, {Geha}, {Gibson}, {Gilmore}, {Glanzman}, {Goodenow},
  {Gressler}, {Gris}, {Guyonnet}, {Hascall}, {Haupt}, {Hernandez}, {Hogan},
  {Huang}, {Huffer}, {Innes}, {Jacoby}, {Jain}, {Jee}, {Jernigan},
  {Jevremovic}, {Johns}, {Jones}, {Juramy-Gilles}, {Juric}, {Kahn}, {Kalirai},
  {Kallivayalil}, {Kalmbach}, {Kantor}, {Kasliwal}, {Kessler}, {Kirkby},
  {Knox}, {Kotov}, {Krabbendam}, {Krughoff}, {Kubanek}, {Kuczewski},
  {Kulkarni}, {Lambert}, {Le Guillou}, {Levine}, {Liang}, {Lim}, {Lintott},
  {Lupton}, {Mahabal}, {Marshall}, {Marshall}, {May}, {McKercher}, {Migliore},
  {Miller}, {Mills}, {Monet}, {Moniez}, {Neill}, {Nief}, {Nomerotski},
  {Nordby}, {O'Connor}, {Oliver}, {Olivier}, {Olsen}, {Ortiz}, {Owen}, {Pain},
  {Peterson}, {Petry}, {Pierfederici}, {Pietrowicz}, {Pike}, {Pinto}, {Plante},
  {Plate}, {Price}, {Prouza}, {Radeka}, {Rajagopal}, {Rasmussen}, {Regnault},
  {Ridgway}, {Ritz}, {Rosing}, {Roucelle}, {Rumore}, {Russo}, {Saha},
  {Sassolas}, {Schalk}, {Schindler}, {Schneider}, {Schumacher}, {Sebag},
  {Sembroski}, {Seppala}, {Shipsey}, {Silvestri}, {Smith}, {Smith}, {Strauss},
  {Stubbs}, {Sweeney}, {Szalay}, {Takacs}, {Thaler}, {Van Berg}, {Vanden Berk},
  {Vetter}, {Virieux}, {Xin}, {Walkowicz}, {Walter}, {Wang}, {Warner},
  {Willman}, {Wittman}, {Wolff}, {Wood-Vasey}, {Yoachim}, {Zhan}, \& {for the
  LSST Collaboration}}]{LSST_1}
{Ivezic}, Z., {Tyson}, J.~A., {Abel}, B., {et~al.} 2008, ArXiv e-prints
  [\eprint[arXiv]{0805.2366}]

\bibitem[{{Jiang} {et~al.}(2014){Jiang}, {Helly}, {Cole}, \&
  {Frenk}}]{Jiang_2014}
{Jiang}, L., {Helly}, J.~C., {Cole}, S., \& {Frenk}, C.~S. 2014, \mnras, 440,
  2115

\bibitem[{{Kaiser}(1987)}]{Kaiser_1987}
{Kaiser}, N. 1987, \mnras, 227, 1

\bibitem[{{Kerscher} {et~al.}(2000){Kerscher}, {Szapudi}, \&
  {Szalay}}]{kerscher}
{Kerscher}, M., {Szapudi}, I., \& {Szalay}, A.~S. 2000, \apj, 535, 13

\bibitem[{{Klypin} \& {Kopylov}(1983)}]{Klypin_1983}
{Klypin}, A.~A. \& {Kopylov}, A.~I. 1983, Soviet Astronomy Letters, 9, 41

\bibitem[{{Kravtsov} {et~al.}(2004){Kravtsov}, {Berlind}, {Wechsler}, {Klypin},
  {Gottl{\"o}ber}, {Allgood}, \& {Primack}}]{Kravtsov_2004_n200}
{Kravtsov}, A.~V., {Berlind}, A.~A., {Wechsler}, R.~H., {et~al.} 2004, \apj,
  609, 35

\bibitem[{{Lagos} {et~al.}(2012){Lagos}, {Bayet}, {Baugh}, {Lacey}, {Bell},
  {Fanidakis}, \& {Geach}}]{lagos}
{Lagos}, C.~d.~P., {Bayet}, E., {Baugh}, C.~M., {et~al.} 2012, \mnras, 426,
  2142

\bibitem[{{Landy} \& {Szalay}(1993)}]{landy_szalay}
{Landy}, S.~D. \& {Szalay}, A.~S. 1993, \apj, 412, 64

\bibitem[{{Laureijs} {et~al.}(2011){Laureijs}, {Amiaux}, {Arduini},
  {Augu{\`e}res}, {Brinchmann}, {Cole}, {Cropper}, {Dabin}, {Duvet}, {Ealet},
  \& et~al.}]{euclid_red_book}
{Laureijs}, R., {Amiaux}, J., {Arduini}, S., {et~al.} 2011, ArXiv e-prints
  [\eprint[arXiv]{1110.3193}]

\bibitem[{{Le F{\`e}vre} {et~al.}(2005){Le F{\`e}vre}, {Vettolani}, {Garilli},
  {Tresse}, {Bottini}, {Le Brun}, {Maccagni}, {Picat}, {Scaramella},
  {Scodeggio}, {Zanichelli}, {Adami}, {Arnaboldi}, {Arnouts}, {Bardelli},
  {Bolzonella}, {Cappi}, {Charlot}, {Ciliegi}, {Contini}, {Foucaud},
  {Franzetti}, {Gavignaud}, {Guzzo}, {Ilbert}, {Iovino}, {McCracken}, {Marano},
  {Marinoni}, {Mathez}, {Mazure}, {Meneux}, {Merighi}, {Paltani}, {Pell{\`o}},
  {Pollo}, {Pozzetti}, {Radovich}, {Zamorani}, {Zucca}, {Bondi}, {Bongiorno},
  {Busarello}, {Lamareille}, {Mellier}, {Merluzzi}, {Ripepi}, \&
  {Rizzo}}]{VVDS}
{Le F{\`e}vre}, O., {Vettolani}, G., {Garilli}, B., {et~al.} 2005, \aap, 439,
  845

\bibitem[{{Lemson} \& {Virgo Consortium}(2006)}]{Lemson_2006}
{Lemson}, G. \& {Virgo Consortium}, t. 2006, ArXiv Astrophysics e-prints
  [\eprint{astro-ph/0608019}]

\bibitem[{Lewis {et~al.}(2000)Lewis, Challinor, \& Lasenby}]{CAMB_1}
Lewis, A., Challinor, A., \& Lasenby, A. 2000, Astrophys. J., 538, 473

\bibitem[{{LSST Science Collaboration} {et~al.}(2009){LSST Science
  Collaboration}, {Abell}, {Allison}, {Anderson}, {Andrew}, {Angel}, {Armus},
  {Arnett}, {Asztalos}, {Axelrod}, \& et~al.}]{LSST_2}
{LSST Science Collaboration}, {Abell}, P.~A., {Allison}, J., {et~al.} 2009,
  ArXiv e-prints [\eprint[arXiv]{0912.0201}]

\bibitem[{{Majumdar} \& {Mohr}(2004)}]{majumdar}
{Majumdar}, S. \& {Mohr}, J.~J. 2004, \apj, 613, 41

\bibitem[{{Marulli} {et~al.}(2012){Marulli}, {Bianchi}, {Branchini}, {Guzzo},
  {Moscardini}, \& {Angulo}}]{Marulli_deprojection}
{Marulli}, F., {Bianchi}, D., {Branchini}, E., {et~al.} 2012, \mnras, 426, 2566

\bibitem[{{Marulli} {et~al.}(2016){Marulli}, {Veropalumbo}, \&
  {Moresco}}]{CosmoBolognaLib}
{Marulli}, F., {Veropalumbo}, A., \& {Moresco}, M. 2016, Astronomy and
  Computing, 14, 35

\bibitem[{{Marulli} {et~al.}(2015){Marulli}, {Veropalumbo}, {Moscardini},
  {Cimatti}, \& {Dolag}}]{Marulli_redshift_evolution}
{Marulli}, F., {Veropalumbo}, A., {Moscardini}, L., {Cimatti}, A., \& {Dolag},
  K. 2015, ArXiv e-prints [\eprint[arXiv]{1505.01170}]

\bibitem[{{Merson} {et~al.}(2013){Merson}, {Baugh}, {Helly}, {Gonzalez-Perez},
  {Cole}, {Bielby}, {Norberg}, {Frenk}, {Benson}, {Bower}, {Lacey}, \&
  {Lagos}}]{alex_merson}
{Merson}, A.~I., {Baugh}, C.~M., {Helly}, J.~C., {et~al.} 2013, \mnras, 429,
  556

\bibitem[{{Mo} \& {White}(1996)}]{Mo_White}
{Mo}, H.~J. \& {White}, S.~D.~M. 1996, \mnras, 282, 347

\bibitem[{{Moscardini} {et~al.}(2000){Moscardini}, {Matarrese}, {Lucchin}, \&
  {Rosati}}]{Moscardini}
{Moscardini}, L., {Matarrese}, S., {Lucchin}, F., \& {Rosati}, P. 2000, \mnras,
  316, 283

\bibitem[{{Norberg} {et~al.}(2011){Norberg}, {Gazta{\~n}aga}, {Baugh}, \&
  {Croton}}]{norberg}
{Norberg}, P., {Gazta{\~n}aga}, E., {Baugh}, C.~M., \& {Croton}, D.~J. 2011,
  \mnras, 418, 2435

\bibitem[{{Peacock} \& {West}(1992)}]{Peacock_west}
{Peacock}, J.~A. \& {West}, M.~J. 1992, \mnras, 259, 494

\bibitem[{{Peebles}(1980)}]{peebles_1980}
{Peebles}, P.~J.~E. 1980, {The large-scale structure of the universe}

\bibitem[{{Rozo} {et~al.}(2009){Rozo}, {Rykoff}, {Koester}, {McKay}, {Hao},
  {Evrard}, {Wechsler}, {Hansen}, {Sheldon}, {Johnston}, {Becker}, {Annis},
  {Bleem}, \& {Scranton}}]{Rozo_2009_n200}
{Rozo}, E., {Rykoff}, E.~S., {Koester}, B.~P., {et~al.} 2009, \apj, 703, 601

\bibitem[{{Rykoff} {et~al.}(2012){Rykoff}, {Koester}, {Rozo}, {Annis},
  {Evrard}, {Hansen}, {Hao}, {Johnston}, {McKay}, \&
  {Wechsler}}]{Rykoff_2012_n200}
{Rykoff}, E.~S., {Koester}, B.~P., {Rozo}, E., {et~al.} 2012, \apj, 746, 178

\bibitem[{{S{\'a}nchez} {et~al.}(2014){S{\'a}nchez}, {Carrasco Kind}, {Lin},
  {Miquel}, {Abdalla}, {Amara}, {Banerji}, {Bonnett}, {Brunner}, {Capozzi},
  {Carnero}, {Castander}, {da Costa}, {Cunha}, {Fausti}, {Gerdes}, {Greisel},
  {Gschwend}, {Hartley}, {Jouvel}, {Lahav}, {Lima}, {Maia}, {Mart{\'{\i}}},
  {Ogando}, {Ostrovski}, {Pellegrini}, {Rau}, {Sadeh}, {Seitz},
  {Sevilla-Noarbe}, {Sypniewski}, {de Vicente}, {Abbot}, {Allam}, {Atlee},
  {Bernstein}, {Bernstein}, {Buckley-Geer}, {Burke}, {Childress}, {Davis},
  {DePoy}, {Dey}, {Desai}, {Diehl}, {Doel}, {Estrada}, {Evrard},
  {Fern{\'a}ndez}, {Finley}, {Flaugher}, {Frieman}, {Gaztanaga}, {Glazebrook},
  {Honscheid}, {Kim}, {Kuehn}, {Kuropatkin}, {Lidman}, {Makler}, {Marshall},
  {Nichol}, {Roodman}, {S{\'a}nchez}, {Santiago}, {Sako}, {Scalzo}, {Smith},
  {Swanson}, {Tarle}, {Thomas}, {Tucker}, {Uddin}, {Vald{\'e}s}, {Walker},
  {Yuan}, \& {Zuntz}}]{Sanchez_DES}
{S{\'a}nchez}, C., {Carrasco Kind}, M., {Lin}, H., {et~al.} 2014, \mnras, 445,
  1482

\bibitem[{{Sartoris} {et~al.}(2016){Sartoris}, {Biviano}, {Fedeli}, {Bartlett},
  {Borgani}, {Costanzi}, {Giocoli}, {Moscardini}, {Weller}, {Ascaso},
  {Bardelli}, {Maurogordato}, \& {Viana}}]{Sartoris_2016}
{Sartoris}, B., {Biviano}, A., {Fedeli}, C., {et~al.} 2016, \mnras, 459, 1764

\bibitem[{{Saunders} {et~al.}(1992){Saunders}, {Rowan-Robinson}, \&
  {Lawrence}}]{saunders_1992}
{Saunders}, W., {Rowan-Robinson}, M., \& {Lawrence}, A. 1992, \mnras, 258, 134

\bibitem[{{Sheth} {et~al.}(2001){Sheth}, {Mo}, \& {Tormen}}]{Sheth}
{Sheth}, R.~K., {Mo}, H.~J., \& {Tormen}, G. 2001, \mnras, 323, 1

\bibitem[{{Spergel} {et~al.}(2007){Spergel}, {Bean}, {Dor{\'e}}, {Nolta},
  {Bennett}, {Dunkley}, {Hinshaw}, {Jarosik}, {Komatsu}, {Page}, {Peiris},
  {Verde}, {Halpern}, {Hill}, {Kogut}, {Limon}, {Meyer}, {Odegard}, {Tucker},
  {Weiland}, {Wollack}, \& {Wright}}]{Spergel_2007}
{Spergel}, D.~N., {Bean}, R., {Dor{\'e}}, O., {et~al.} 2007, \apjs, 170, 377

\bibitem[{{Springel}(2005)}]{Springel_from_Merson}
{Springel}, V. 2005, \mnras, 364, 1105

\bibitem[{{Springel} {et~al.}(2005){Springel}, {White}, {Jenkins}, {Frenk},
  {Yoshida}, {Gao}, {Navarro}, {Thacker}, {Croton}, {Helly}, {Peacock}, {Cole},
  {Thomas}, {Couchman}, {Evrard}, {Colberg}, \& {Pearce}}]{springel}
{Springel}, V., {White}, S.~D.~M., {Jenkins}, A., {et~al.} 2005, \nat, 435, 629

\bibitem[{{Taylor}(2005)}]{topcat}
{Taylor}, M.~B. 2005, in Astronomical Society of the Pacific Conference Series,
  Vol. 347, Astronomical Data Analysis Software and Systems XIV, ed.
  P.~{Shopbell}, M.~{Britton}, \& R.~{Ebert}, 29

\bibitem[{{Tinker} {et~al.}(2010){Tinker}, {Robertson}, {Kravtsov}, {Klypin},
  {Warren}, {Yepes}, \& {Gottl{\"o}ber}}]{Tinker_2010}
{Tinker}, J.~L., {Robertson}, B.~E., {Kravtsov}, A.~V., {et~al.} 2010, \apj,
  724, 878

\bibitem[{{Totsuji} \& {Kihara}(1969)}]{totsuji_1969}
{Totsuji}, H. \& {Kihara}, T. 1969, \pasj, 21, 221

\bibitem[{{Veropalumbo} {et~al.}(2014){Veropalumbo}, {Marulli}, {Moscardini},
  {Moresco}, \& {Cimatti}}]{Veropalumbo_1}
{Veropalumbo}, A., {Marulli}, F., {Moscardini}, L., {Moresco}, M., \&
  {Cimatti}, A. 2014, \mnras, 442, 3275

\bibitem[{{Veropalumbo} {et~al.}(2016){Veropalumbo}, {Marulli}, {Moscardini},
  {Moresco}, \& {Cimatti}}]{Veropalumbo_2}
{Veropalumbo}, A., {Marulli}, F., {Moscardini}, L., {Moresco}, M., \&
  {Cimatti}, A. 2016, \mnras, 458, 1909

\bibitem[{Younger {et~al.}(2005)Younger, Bahcall, \& Bode}]{Younger}
Younger, J.~D., Bahcall, N.~A., \& Bode, P. 2005, Astrophys. J., 622, 1

\bibitem[{{Zehavi} {et~al.}(2005){Zehavi}, {Zheng}, {Weinberg}, {Frieman},
  {Berlind}, {Blanton}, {Scoccimarro}, {Sheth}, {Strauss}, {Kayo}, {Suto},
  {Fukugita}, {Nakamura}, {Bahcall}, {Brinkmann}, {Gunn}, {Hennessy},
  {Ivezi{\'c}}, {Knapp}, {Loveday}, {Meiksin}, {Schlegel}, {Schneider},
  {Szapudi}, {Tegmark}, {Vogeley}, {York}, \& {SDSS Collaboration}}]{zehavi_b}
{Zehavi}, I., {Zheng}, Z., {Weinberg}, D.~H., {et~al.} 2005, \apj, 630, 1

\bibitem[{{Zheng} {et~al.}(2005){Zheng}, {Berlind}, {Weinberg}, {Benson},
  {Baugh}, {Cole}, {Dav{\'e}}, {Frenk}, {Katz}, \& {Lacey}}]{Zheng_2005_n200}
{Zheng}, Z., {Berlind}, A.~A., {Weinberg}, D.~H., {et~al.} 2005, \apj, 633, 791

\end{thebibliography}


\appendix

\section{Values of best-fit parameters from the two-point correlation fit and bias values for photometric redshift catalogues}

Table \ref{table:r0_vs_gamma_zcosmo_vs_zphot} shows the values of the best-fit 
parameters (as shown in Figure 
\ref{fig:z_vs_r0_zcosmo_vs_zphot}) for the two-point correlation 
function of the four sub-samples with redshift errors 
$\sigma_z/(1+z_{c}) = 0.005,0.010,0.030$, and $0.050$, and the corresponding
values obtained for the parent catalogue with cosmological redshift ($z_c$). 

For the same four sub-samples and the parent sample  with 
cosmological redshift ($z_c$), 
tables \ref{table:bias_values_zphot_1_and_2} and 
\ref{table:bias_values_zphot_3_and_4} show the bias values 
(see also Figure \ref{fig:z_vs_bias_zcosmo_vs_zphot}).

\begin{table*}
\centering
\caption{Number of clusters in a given redshift range for $z_{c}$ and $z_{phot}$ with mass cut 
$M_{halo}>5\times 10^{13}\;h^{-1}\;M_{\odot}$. 
The $z_{phot}$ uncertainties are $\sigma_z/(1+z_{c}) = 0.005,0.010,0.030$, and $0.050$. 
(1) Redshift range, (2) number of clusters in $z_{c}$ window, 
(3) number of clusters in $z_{phot}$ window, (4) common clusters, (5) uncommon clusters and (6) the percentage
of contaminants are quoted.}
\label{table:contaminants_table}
\begin{tabular}{cccccc}
\hline
Redshift range & 
\multicolumn{1}{p{2cm}}{\centering Clusters in \\ $z_{c}$ window } & 
\multicolumn{1}{p{3cm}}{\centering Clusters in \\ $z_{phot}$ window } & 
Common & Uncommon & \% contaminants  \\
\hline 

\multicolumn{6}{c}{\textbf{$\sigma_z = 0.005\times(1+z_{c})$}} \\
\hline 
\rule{0pt}{2.5ex}
0.1$<z<$0.4 & 3210 & 3214 & 3160 & 50 & 1.55 \\

\rule{0pt}{2.5ex}
0.4$<z<$0.7 & 7301 & 7310 & 7162 & 139 & 1.90 \\
 
\rule{0pt}{2.5ex}
0.7$<z<$1.0 & 8128 & 8088 & 7933 & 195 & 2.39 \\
 
\rule{0pt}{2.5ex}
1.0$<z<$1.3 & 5963 & 6001 & 5842 & 121 & 2.02 \\

\rule{0pt}{2.5ex}
1.3$<z<$1.6 & 3365 & 3356 & 3252 & 113 & 3.35 \\

\rule{0pt}{2.5ex}
1.6$<z<$2.1 & 2258 & 2251 & 2197 & 61 & 2.70 \\
\hline 

\multicolumn{6}{c}{\textbf{$\sigma_z = 0.010\times(1+z_{c})$}} \\ 
\hline
\rule{0pt}{2.5ex}
0.1$<z<$0.4 & 3210 & 3216 & 3115 & 95 & 2.95 \\

\rule{0pt}{2.5ex}
0.4$<z<$0.7 & 7301 & 7338 & 7042 & 259 & 3.54 \\

\rule{0pt}{2.5ex}
0.7$<z<$1.0 & 8128 & 8095 & 7745 & 383 & 4.71 \\

\rule{0pt}{2.5ex}
1.0$<z<$1.3 & 5963 & 5973 & 5676 & 287 & 4.81 \\
\rule{0pt}{2.5ex}
1.3$<z<$1.6 & 3365 & 3350 & 3144 & 221 & 6.56 \\
\rule{0pt}{2.5ex}
1.6$<z<$2.1 & 2258 & 2239 & 2133 & 125 & 5.53 \\
\hline 
\multicolumn{6}{c}{\textbf{$\sigma_z = 0.030\times(1+z_{c})$}} \\ 
\hline \rule{0pt}{2.5ex}
0.1$<z<$0.4 & 3210 & 3196 & 2884 & 326 & 10.15 \\
\rule{0pt}{2.5ex}
0.4$<z<$0.7 & 7301 & 7396 & 6492 & 809 & 11.08 \\
\rule{0pt}{2.5ex}
0.7$<z<$1.0 & 8128 & 8053 & 6972 & 1156 & 14.22 \\
\rule{0pt}{2.5ex}
1.0$<z<$1.3 & 5963 & 5880 & 4958 & 1005 & 16.85 \\
\rule{0pt}{2.5ex}
1.3$<z<$1.6 & 3365 & 3388 & 2712 & 653 & 19.40 \\
\rule{0pt}{2.5ex}
1.6$<z<$2.1 & 2258 & 2251 & 1922 & 336 & 14.88 \\
\hline 
\multicolumn{6}{c}{\textbf{$\sigma_z = 0.050\times(1+z_{c})$}} \\ 
\hline
\rule{0pt}{2.5ex}
0.1$<z<$0.4 & 3210 & 3205 & 2647 & 563 & 17.53 \\
\rule{0pt}{2.5ex}
0.4$<z<$0.7 & 7301 & 7433 & 5906 & 1395 & 19.10 \\
\rule{0pt}{2.5ex}
0.7$<z<$1.0 & 8128 & 7937 & 6153 & 1975 & 24.29 \\
\rule{0pt}{2.5ex}
1.0$<z<$1.3 & 5963 & 5859 & 4294 & 1669 & 27.98 \\
\rule{0pt}{2.5ex}
1.3$<z<$1.6 & 3365 & 3352 & 2248 & 1117 & 33.19 \\
\rule{0pt}{2.5ex}
1.6$<z<$2.1 & 2258 & 2277 & 1717 & 541 & 23.95 \\
\hline 

\end{tabular} 
\end{table*}

\begin{table*}
\centering
\caption{Parameters obtained from the fit for the real-space correlation function \xir{} on the ideal 
zero-error simulation for the different redshift cut catalogues and the same obtained from the photometric 
redshift catalogues with $\sigma_z/(1+z_{c}) = 0.005,0.010,0.030$, and $0.050$. (1) Redshift cut used, (2) 
photometric uncertainty
$\sigma_z/(1+z_{c})$, (3) correlation length $r_{0}$, (4) slope $\gamma$ and (5) the number of 
clusters $N_{clusters}$,
(6) median redshift, and (7) mean redshift.} 
\label{table:r0_vs_gamma_zcosmo_vs_zphot}
\begin{tabular}{*{7}{c}} 
\toprule
\multicolumn{1}{p{2cm}}{Redshift range}
& \multicolumn{1}{p{2cm}}{\centering $\sigma_z/(1+z_{c})$}
& \multicolumn{1}{p{2cm}}{\centering $r_{0}\;h^{-1}$Mpc} & 
$\gamma$ & $N_{clusters}$ & Median $z$ & Mean $z$ \\ 
\midrule
\multirow{5}{*}{$0.1<z<0.4$} & $z_c$ & 12.22$\pm$0.26 & 1.90$\pm$0.05
			 	 & 3210 & 0.30 & 0.29 \\ [1ex] 
   		& 0.005 & 11.99$\pm$0.44 & 2.01$\pm$0.09  & 3214 & 0.30 & 0.29 \\ [1ex] 
   		& 0.010 & 12.47$\pm$0.51 & 1.80$\pm$0.10 & 3216 & 0.30 & 0.29 \\ [1ex] 
   		& 0.030 & 12.57$\pm$0.64 & 1.97$\pm$0.13 & 3196 & 0.30 & 0.28 \\ [1ex] 
   		& 0.050 & 12.48$\pm$0.61 & 1.99$\pm$0.12 & 3205 & 0.29 & 0.28 \\ [1ex] \hline \noalign{\vskip 0.1cm} 

\multirow{5}{*}{$0.4<z<0.7$} & $z_c$ & 13.20$\pm$0.23 & 1.98$\pm$0.05
			 	 & 7301 & 0.56 & 0.55 \\ [1ex] 
   		& 0.005 & 12.89$\pm$0.26 & 1.94$\pm$0.06  & 7310 & 0.56 & 0.55  \\ [1ex] 
   		& 0.010 & 12.84$\pm$0.63 & 1.93$\pm$0.08 & 7338 & 0.56 & 0.56 \\ [1ex] 
   		& 0.030 & 12.91$\pm$0.72 & 2.02$\pm$0.12 & 7396 & 0.56 & 0.55 \\ [1ex] 
   		& 0.050 & 12.88$\pm$0.76 & 1.90$\pm$0.14 & 7433 & 0.56 & 0.55 \\ [1ex] \hline \noalign{\vskip 0.1cm}   
   		
\multirow{5}{*}{$0.7<z<1.0$} & $z_c$ & 14.86$\pm$0.33 & 1.97$\pm$0.05
			 	 & 8128 & 0.84 & 0.84 \\ [1ex] 
   		& 0.005 & 15.07$\pm$0.49 & 1.87$\pm$0.07  & 8088 & 0.84 & 0.84  \\ [1ex] 
   		& 0.010 & 14.99$\pm$1.00 & 1.90$\pm$0.18 & 8095 & 0.84 & 0.84  \\ [1ex] 
   		& 0.030 & 14.06$\pm$0.63 & 2.05$\pm$0.19 & 8053 & 0.84 & 0.84 \\ [1ex] 
   		& 0.050 & 14.36$\pm$0.83 & 1.91$\pm$0.16 & 7937 & 0.84 & 0.84 \\ [1ex] \hline \noalign{\vskip 0.1cm}   
   		
\multirow{5}{*}{$1.0<z<1.3$} & $z_c$ & 17.00$\pm$0.48 & 2.00$\pm$0.07
			 	 & 5963 & 1.13 & 1.13 \\ [1ex] 
   		& 0.005 & 17.29$\pm$0.64 & 2.06$\pm$0.08  & 6001 & 1.13 & 1.13  \\ [1ex] 
   		& 0.010 & 16.90$\pm$0.87 & 2.08$\pm$0.12 & 5973 & 1.13 & 1.14 \\ [1ex] 
   		& 0.030 & 17.45$\pm$1.14 & 1.88$\pm$0.16 & 5880 & 1.13 & 1.14 \\ [1ex] 
   		& 0.050 & 17.43$\pm$0.92 & 1.90$\pm$0.13 & 5859 & 1.13 & 1.14 \\ [1ex] \hline \noalign{\vskip 0.1cm}

\multirow{5}{*}{$1.3<z<1.6$} & $z_c$ & 18.26$\pm$0.62 & 2.15$\pm$0.06
			 	 & 3365 & 1.43 & 1.43 \\ [1ex] 
   		& 0.005 & 18.31$\pm$0.51 & 2.09$\pm$0.08  & 3356 & 1.43 & 1.43  \\ [1ex] 
   		& 0.010 & 18.31$\pm$0.75 & 2.26$\pm$0.21 & 3350 & 1.43 & 1.43 \\ [1ex] 
   		& 0.030 & 19.22$\pm$1.16 & 2.16$\pm$0.17 & 3388 & 1.42 & 1.43 \\ [1ex] 
   		& 0.050 & 18.78$\pm$1.35 & 2.11$\pm$0.18 & 3352 & 1.43 & 1.43 \\ [1ex] \hline \noalign{\vskip 0.1cm}

\multirow{5}{*}{$1.6<z<2.1$} & $z_c$ & 19.48$\pm$1.41 & 2.23$\pm$0.21
			 	 & 2258 & 1.76 & 1.79 \\ [1ex] 
   		& 0.005 & 18.86$\pm$0.88 & 2.16$\pm$0.23  & 2251 & 1.77 & 1.79  \\ [1ex] 
   		& 0.010 & 18.76$\pm$0.85 & 2.32$\pm$0.27 & 2239 & 1.77 & 1.79 \\ [1ex] 
   		& 0.030 & 18.74$\pm$1.89 & 2.15$\pm$0.17 & 2251 & 1.77 & 1.79 \\ [1ex] 
   		& 0.050 & 19.62$\pm$2.16 & 2.15$\pm$0.26 & 2277 & 1.77 & 1.79 \\ [1ex] \hline

\end{tabular}
\end{table*}

\begin{table*}[t]
\centering
\begin{tabular}{|c|cccc|} 
\hline
\rule{0pt}{2.5ex}
$\sigma_z/(1+z_{c})$ & 
\multicolumn{1}{p{2cm}}{\centering Mass ($h^{-1}\;M_{\odot}$)} & 
Redshift range & Bias & $N_{clusters}$ \\ 
\hline
\rule{0pt}{2.5ex}
\multirow{18}{*}{0.005} & \multirow{5}{*}{$2\times10^{13}$} & $0.1<z<0.4$ & 1.71$\pm$0.13 & 10521 \\ [1ex]
													&	& $0.4<z<0.7$ & 1.97$\pm$0.06 & 27224 \\ [1ex] 
													&	& $0.7<z<1.0$ & 2.46$\pm$0.06 & 35045 \\ [1ex] 
													&	& $1.0<z<1.3$ & 2.86$\pm$0.23 & 31845 \\ [1ex]
													&	& $1.3<z<1.6$ & 3.31$\pm$0.30 & 23017 \\ [1ex]
													&	& $1.6<z<2.1$ & 4.23$\pm$0.57 & 18904 \\ [1ex]
													
\cline{2-5}
\rule{0pt}{2.5ex} 
					   & \multirow{5}{*}{$5\times10^{13}$} & $0.1<z<0.4$ & 2.17$\pm$0.23 & 3214 \\ [1ex]
													&	& $0.4<z<0.7$ & 2.78$\pm$0.19 & 7310 \\ [1ex] 
													&	& $0.7<z<1.0$ & 3.34$\pm$0.26 & 8088 \\ [1ex] 
													&	& $1.0<z<1.3$ & 4.17$\pm$0.06 & 6001 \\ [1ex]
													&	& $1.3<z<1.6$ & 6.10$\pm$0.81 & 3356 \\ [1ex]
													&	& $1.6<z<2.1$ & 5.75$\pm$0.81 & 2251 \\ [1ex]
\cline{2-5}
\rule{0pt}{2.5ex}
					   & \multirow{5}{*}{$1\times10^{14}$} & $0.1<z<0.4$ & 2.72$\pm$0.27 & 1116 \\ [1ex]
													&	& $0.4<z<0.7$ & 3.86$\pm$0.37 & 2231 \\ [1ex] 
													&	& $0.7<z<1.0$ & 5.07$\pm$0.41 & 2065 \\ [1ex] 
													&	& $1.0<z<1.3$ & 6.35$\pm$1.29 & 1218 \\ [1ex]
													&	& $1.3<z<1.6$ & 8.12$\pm$0.98 & 594 \\ [1ex]
\cline{2-5} 
\rule{0pt}{2.5ex}
					   & \multirow{3}{*}{$2\times10^{14}$} & $0.1<z<0.4$ & 3.78$\pm$1.48 & 316 \\ [1ex]
													&	& $0.4<z<0.7$ & 4.88$\pm$1.07 & 544 \\ [1ex] 
													&	& $0.7<z<1.0$ & 5.45$\pm$1.38 & 399 \\ [1ex] 
\hline
\rule{0pt}{2.5ex}
\multirow{18}{*}{0.010} & \multirow{5}{*}{$2\times10^{13}$} & $0.1<z<0.4$ & 1.76$\pm$0.08 & 10536 \\ [1ex]
													&	& $0.4<z<0.7$ & 1.91$\pm$0.14 & 27283 \\ [1ex] 
													&	& $0.7<z<1.0$ & 2.48$\pm$0.09 & 35022 \\ [1ex] 
													&	& $1.0<z<1.3$ & 2.96$\pm$0.19 & 31763 \\ [1ex]
													&	& $1.3<z<1.6$ & 3.39$\pm$0.32 & 23021 \\ [1ex]
													&	& $1.6<z<2.1$ & 4.10$\pm$0.47 & 18898 \\ [1ex]
\cline{2-5}
\rule{0pt}{2.5ex}
					   & \multirow{5}{*}{$5\times10^{13}$} & $0.1<z<0.4$ & 2.23$\pm$0.12 & 3216 \\ [1ex]
													&	& $0.4<z<0.7$ & 2.80$\pm$0.13 & 7338 \\ [1ex] 
													&	& $0.7<z<1.0$ & 3.31$\pm$0.37 & 8095 \\ [1ex] 
													&	& $1.0<z<1.3$ & 4.07$\pm$0.59 & 5973 \\ [1ex]
													&	& $1.3<z<1.6$ & 5.76$\pm$0.56 & 3350 \\ [1ex]
													&	& $1.6<z<2.1$ & 6.01$\pm$0.96 & 2239 \\ [1ex]
\cline{2-5}
\rule{0pt}{2.5ex}
					   & \multirow{5}{*}{$1\times10^{14}$} & $0.1<z<0.4$ & 2.86$\pm$0.33 & 1124 \\ [1ex]
													&	& $0.4<z<0.7$ & 3.58$\pm$0.53 & 2235 \\ [1ex] 
													&	& $0.7<z<1.0$ & 4.96$\pm$0.75 & 2069 \\ [1ex] 
													&	& $1.0<z<1.3$ & 5.57$\pm$2.36 & 1210 \\ [1ex]
													&	& $1.3<z<1.6$ & 7.81$\pm$3.12 & 587 \\ [1ex]
\cline{2-5} 
\rule{0pt}{2.5ex}
					   & \multirow{3}{*}{$2\times10^{14}$} & $0.1<z<0.4$ & 4.56$\pm$0.74 & 318 \\ [1ex]
													&	& $0.4<z<0.7$ & 5.47$\pm$2.3 & 547 \\ [1ex] 
													&	& $0.7<z<1.0$ & 7.16$\pm$1.31 & 394 \\ [1ex] 

\hline 
\end{tabular}
\caption{Bias values obtained for the first two photometric redshift catalogues 
($\sigma_z/(1+z_{c}) = 0.005$ and $0.010$) with the four mass threshold cuts in the five redshift bins used. 
(1) Photometric uncertainty $\sigma_z/(1+z_{c})$, (2) mass cut $M_{halo}$ cut, (3) redshift range, (4) 
the bias, and (5) the number of clusters $N_{clusters}$ are given. } 
\label{table:bias_values_zphot_1_and_2}
\end{table*}

\begin{table*}[t]
\centering
\begin{tabular}{|c|cccc|} 
\hline
\rule{0pt}{2.5ex}
$\sigma_z/(1+z_{c})$ & 
\multicolumn{1}{p{2cm}}{\centering Mass ($h^{-1}\;M_{\odot}$)} & 
Redshift range & Bias & $N_{clusters}$ \\ 
\hline
\rule{0pt}{2.5ex}
\multirow{18}{*}{0.030} & \multirow{5}{*}{$2\times10^{13}$} & $0.1<z<0.4$ & 1.77$\pm$0.15 & 10581 \\ [1ex]
													&	& $0.4<z<0.7$ & 2.05$\pm$0.14 & 27475 \\ [1ex] 
													&	& $0.7<z<1.0$ & 2.36$\pm$0.20 & 34849 \\ [1ex] 
													&	& $1.0<z<1.3$ & 3.13$\pm$0.36 & 31457 \\ [1ex]
													&	& $1.3<z<1.6$ & 3.67$\pm$0.49 & 23028 \\ [1ex]
													&	& $1.6<z<2.1$ & 5.14$\pm$0.38 & 18863 \\ [1ex]
\cline{2-5}
\rule{0pt}{2.5ex}
					   & \multirow{5}{*}{$5\times10^{13}$} & $0.1<z<0.4$ & 2.38$\pm$0.45 & 3196 \\ [1ex]
													&	& $0.4<z<0.7$ & 2.97$\pm$0.07 & 7396 \\ [1ex] 
													&	& $0.7<z<1.0$ & 3.48$\pm$0.51 & 8053 \\ [1ex] 
													&	& $1.0<z<1.3$ & 4.33$\pm$0.83 & 5880 \\ [1ex]
													&	& $1.3<z<1.6$ & 4.99$\pm$1.02 & 3388 \\ [1ex]
													&	& $1.6<z<2.1$ & 6.66$\pm$0.93 & 2251 \\ [1ex]
\cline{2-5}
\rule{0pt}{2.5ex}
					   & \multirow{5}{*}{$1\times10^{14}$} & $0.1<z<0.4$ & 2.83$\pm$0.43 & 1115 \\ [1ex]
													&	& $0.4<z<0.7$ & 3.70$\pm$0.49 & 2253 \\ [1ex] 
													&	& $0.7<z<1.0$ & 4.63$\pm$2.05 & 2041 \\ [1ex] 
													&	& $1.0<z<1.3$ & 5.93$\pm$2.24 & 1228 \\ [1ex]
													&	& $1.3<z<1.6$ & 7.54$\pm$1.74 &580  \\ [1ex]
\cline{2-5} 
\rule{0pt}{2.5ex}
					   & \multirow{3}{*}{$2\times10^{14}$} & $0.1<z<0.4$ & 4.74$\pm$0.61 & 317 \\ [1ex]
													&	& $0.4<z<0.7$ & 5.37$\pm$1.50 & 554 \\ [1ex] 
													&	& $0.7<z<1.0$ & 6.53$\pm$2.30 & 383 \\ [1ex] 
\hline
\rule{0pt}{2.5ex}
\multirow{18}{*}{0.050} & \multirow{5}{*}{$2\times10^{13}$} & $0.1<z<0.4$ & 1.71$\pm$0.28 & 10835 \\ [1ex]
													&	& $0.4<z<0.7$ & 2.05$\pm$0.19 & 27570 \\ [1ex] 
													&	& $0.7<z<1.0$ & 2.71$\pm$0.28 & 34575 \\ [1ex] 
													&	& $1.0<z<1.3$ & 3.34$\pm$0.29 & 31168 \\ [1ex]
													&	& $1.3<z<1.6$ & 4.20$\pm$0.86 & 22733 \\ [1ex]
													&	& $1.6<z<2.1$ & 5.01$\pm$0.63 & 18889 \\ [1ex]
\cline{2-5}
\rule{0pt}{2.5ex}
					   & \multirow{5}{*}{$5\times10^{13}$} & $0.1<z<0.4$ & 2.40$\pm$0.18 & 3205 \\ [1ex]
													&	& $0.4<z<0.7$ & 2.62$\pm$0.21 & 7443 \\ [1ex] 
													&	& $0.7<z<1.0$ & 3.06$\pm$0.29 & 7937 \\ [1ex] 
													&	& $1.0<z<1.3$ & 4.21$\pm$0.24 & 5859 \\ [1ex]
													&	& $1.3<z<1.6$ & 6.31$\pm$1.01 & 3352 \\ [1ex]
													&	& $1.6<z<2.1$ & 7.02$\pm$0.63 & 2277 \\ [1ex]
\cline{2-5}
\rule{0pt}{2.5ex}
					   & \multirow{5}{*}{$1\times10^{14}$} & $0.1<z<0.4$ & 2.98$\pm$0.61 & 1109 \\ [1ex]
													&	& $0.4<z<0.7$ & 3.37$\pm$1.20 & 2258 \\ [1ex] 
													&	& $0.7<z<1.0$ & 4.89$\pm$0.86 & 1992 \\ [1ex] 
													&	& $1.0<z<1.3$ & 5.75$\pm$2.30 & 1257 \\ [1ex]
													&	& $1.3<z<1.6$ & 9.75$\pm$2.79 & 562 \\ [1ex]
\cline{2-5} 
\rule{0pt}{2.5ex}
					   & \multirow{3}{*}{$2\times10^{14}$} & $0.1<z<0.4$ & 5.12$\pm$1.08 & 311 \\ [1ex]
													&	& $0.4<z<0.7$ & 5.57$\pm$2.40 & 560 \\ [1ex] 
													&	& $0.7<z<1.0$ & 6.91$\pm$2.97 & 370 \\ [1ex] 

\hline 
\end{tabular}
\caption{Bias values obtained for the last two photometric redshift catalogues ($\sigma_z/(1+z_{c}) = 0.030$ 
and $0.050$) with the four mass threshold cuts in the five redshift bins used. 
(1) Photometric uncertainty $\sigma_z/(1+z_{c})$, 
(2) mass cut $M_{halo}$ cut, (3) redshift range, (4) the bias, and (5) 
the number of clusters $N_{clusters}$ are given.  } 
\label{table:bias_values_zphot_3_and_4}
\end{table*}

%
%

\end{document}